\documentclass[english,aps,prl,reprint,superscriptaddress,floatfix,amsfonts]{revtex4-1}
\usepackage[latin9]{inputenc}
\usepackage{color}
\usepackage{bm}
\usepackage{multirow}
\usepackage{amsmath}
\usepackage{graphicx}
\usepackage{amssymb}
\usepackage{amsfonts}
\usepackage{bbold}

\makeatletter

\usepackage{epstopdf}
\usepackage{dcolumn}
\usepackage{bm}
\usepackage{babel}
\usepackage{multirow}

\makeatother

\usepackage{babel}

\begin{document}


\title{Effective momentum-momentum coupling in a correlated electronic system}


\author{T. V. Trevisan}
\altaffiliation[Present address: ]{Ames Laboratory, Ames, Iowa 50011, USA.}
\affiliation{Instituto de F{\'i}sica Gleb Wataghin, Universidade Estadual de Campinas, UNICAMP, 13083-859 Campinas, S{\~a}o Paulo, Brazil}
\author{G. M. Monteiro}
\altaffiliation[Present address: ]{Department of Physics, City College, City University of New York, New York, NY 10031, USA.}
\affiliation{Instituto de F{\'i}sica Gleb Wataghin, Universidade Estadual de Campinas, UNICAMP, 13083-859 Campinas, S{\~a}o Paulo, Brazil}
\author{A. O.  Caldeira}
\affiliation{Instituto de F{\'i}sica Gleb Wataghin, Universidade Estadual de Campinas, UNICAMP, 13083-859 Campinas, S{\~a}o Paulo, Brazil}


\date{\today}

\begin{abstract}
We present a way of partly reincorporate the effects of the localized bonding electrons on the dynamics of their itinerant counterparts in Hubbard-like Hamiltonians. This is done by relaxing the constraint that the former should be entirely frozen in the chemical bonds between the underlying lattice sites through the employment of a Born-Oppenheimer-like ansatz for the wavefunction of the whole electronic system.  Accordingly, the latter includes itinerant as well as bonding electron coordinates. Going beyond the adiabatic approximation, we show that the net effect of virtual transitions of bonding electrons between their ground and excited states is to furnish the itinerant electrons with an effective inter-electronic momentum-momentum interaction.  Although we have applied these ideas to the specific case of rings, our assumptions can be generalized to higher dimensional systems sharing the required properties of which we have made use herein. 
\end{abstract}


\maketitle


\section{Introduction\label{Intro}}

The existence of strongly interacting many-body systems is the general rule rather than an exception. Everything we know of is composed of many interacting parts, and the realization that the strength of the interaction between these parts may vary within a vast range allows the separation of length, time and energy scales. Such scales separation often helps to establish criteria by which the dynamics of the systems are greatly simplified. This does not mean that their study can be reduced to triviality, but rather creates a hierarchy of different levels of complexity, even having neglected a huge number of physical effects which play a minor role with respect to our aforementioned criteria.

There is a plethora of examples following this modeling scheme, but let us restrict ourselves to a few of those which deal with \textit{electronic systems}. The reason behind this choice is twofold. Firstly, the unquestionable relevance of electrons for the understanding of atomic and molecular structure, chemical reactions, electric and magnetic properties of solids, or, more generally, condensed matter systems, and the challenging physics of nano and mesoscopic devices. Secondly, electrons under some specific conditions are the entities we shall be addressing in this work.

If we focus on the low energy physics of a many-electron system, we know that the Coulomb interaction is all there is. Electron-electron, ion-ion, or electron-ion interactions can all ultimately be described by Coulomb forces. Relativistic corrections include spin-orbit, spin-spin (magnetic dipole), and current-current (Breit-Darwin) interactions \cite{Salpeter,Breit1,Darwin}. All of them are of electromagnetic origin and have their importance dictated by the band structure, surface effects, and/or many other specific constraints to which the system under investigation might be subject. So, in principle, there is no secret about the basic interactions underlying a many-electron system. Nevertheless, even with the advent of very fast and powerful computers, a full understanding of these systems is known to be absolutely out of reach. To make use of these computational resources to get information about the system, one still has to appeal to a set of ingenious numerical or simulation methods. 

The computational approach to the approximate description of many-electron systems is undeniably of fundamental importance. Still, it becomes particularly more useful when complemented with some input from simpler models from which more physical insight can be extracted. These models are the simplified versions of the realistic situations we have mentioned above, and they are meant to capture the relevant physics of the electronic system under consideration. It must be emphasized that, even for these simplified models, exact solutions are only rarely accessible.

A successful example is the Landau's Fermi liquid theory \cite{landau1}, which replaces strongly interacting neutral fermions (liquid $^{3}$He, for example) by weakly coupled fermionic quasi-particles with renormalized properties. Importantly, the Fermi liquid theory was also extended to treat charged fermions \cite{pinesQL}, and became a paradigm for the theory of interacting electrons in metals. Although this theory was initially proposed only on phenomenological grounds, Landau himself \cite{landau2} put forward a more microscopic justification for his own model. The Fermi Liquid theory, together with the collective description of interactions in an electron gas, constitute the basic approach for the theory of many-electron systems. 

Other examples involves metallic crystals, which can be modeled in two particularly simple ways \cite{AM}: the nearly free electron and the tight-binding models. In both cases, the electrons are considered independent (except for their fermionic statistics). This hypothesis is supported by the strong screening effects, which drastically reduces the Coulomb interaction between them, and is particularly important for very dense electronic systems. In more dilute systems, the effect of the undressed Coulomb interaction is more pronounced. Importantly, interactions in metallic crystals can be accounted for either by the Hartree-Fock method \cite{AM, Fetter}, in the weakly bound case, or by the Hubbard model\cite{Hubbard1} in the tight-binding case.

In both the nearly free electron model and the tight-binding model, the lattice potential is considered static, which is enough to have a good approximate description of the electronic band structure of the system. However, to study transport phenomena \cite{mahan,Rammer}, imperfections in the periodic lattice and/or its own dynamics matter for the calculation of response functions such as, for instance, the electric conductivity. Besides, lattice dynamical fluctuations are known to mediate an effective inter-electronic attractive interaction, which might result in the formation of Cooper pairs \cite{cooper}, the leading charge carriers in the theory of superconductivity\cite{bcs}. Effective inter-electronic interactions are quite common in condensed matter systems and may be mediated by different excitations or components of the medium \cite{piers}.

In this work, we shall discuss the appearance of an effective inter-electronic interaction mediated by the electrons themselves. We argue that there is a natural separation between the energy scales of the bonding    and the itinerant electrons. Relaxing the often assumed condition that the bonding electrons are frozen in the bonds, we explicitly show that their virtual excitations induce a momentum-momentum coupling between the itinerant electrons. Although we develop our model for some specific systems, namely small rings, we argue that the physical reasoning that led us to the finding reported here can be extended to many other systems.

This paper is organized as follows: in Sec.\ref{Model}, to set the notation we use throughout this manuscript, as well as to make it self-contained, we briefly discuss the essential features of the standard single-band Hubbard model accounting solely for the degrees of freedom of the itinerant electrons. In Sec.\ref{H0new}, we relax the constraint that the bonding electrons must be frozen in the chemical bonds, and argue that an appropriate way to cope with the resulting physics involves going beyond the well-known Born-Oppenheimer approximation to treat the coupling between itinerant and bonding electrons. This procedure is shown to generate an effective momentum-momentum coupling between the itinerant electrons. The derivation of such effective interaction in first- and second-quantization is shown in Secs.\ref{first} and \ref{W2Q}, respectively. Intermediate steps crucial for the derivation of the expressions of the main text are presented in Appendices \ref{P} and \ref{w}. We summarize our results in Sec.\ref{C}. 

\section{One-dimensional Hubbard rings\label{Model}}

We study electrons in small discrete rings, i.e., electrons in $1D$ lattices with periodic boundary conditions when it has a finite and small number of sites. We denote by $N$ the number of sites of the ring and $a$ its lattice spacing so that the ring length is $L=Na$. These discrete rings are sometimes called \textit{Hubbard rings}\cite{Maiti} if their electronic degrees of freedom are modeled by the Hubbard model \cite{Hubbard1} or some extension thereof, as is the case in this work. In the limit $N\rightarrow\infty, a\rightarrow\,\textrm{0},\, \textrm{but finite}\,L$, there are many different approaches to describe either approximate \cite{doniach} or even exact solutions \cite{lw} for the electronic problem whereas for finite, but small $N$, exact diagonalization is always an accessible way out of this problem. 

The Hubbard model considers, on top of the on-site repulsion between itinerant electrons, their interaction with the effective lattice potential composed by the lattice ionic potential together with the one created by the assumed frozen cloud of bonding electrons. Our goal here is to analyze the role played by some of the latter in this specific electronic problem. We revisit some of the essential steps taken when obtaining the Hubbard model and establish the conditions under which they will dynamically affect the model, and modify its original form.

\begin{figure}[t!]
\centering
\includegraphics[width=0.9\linewidth]{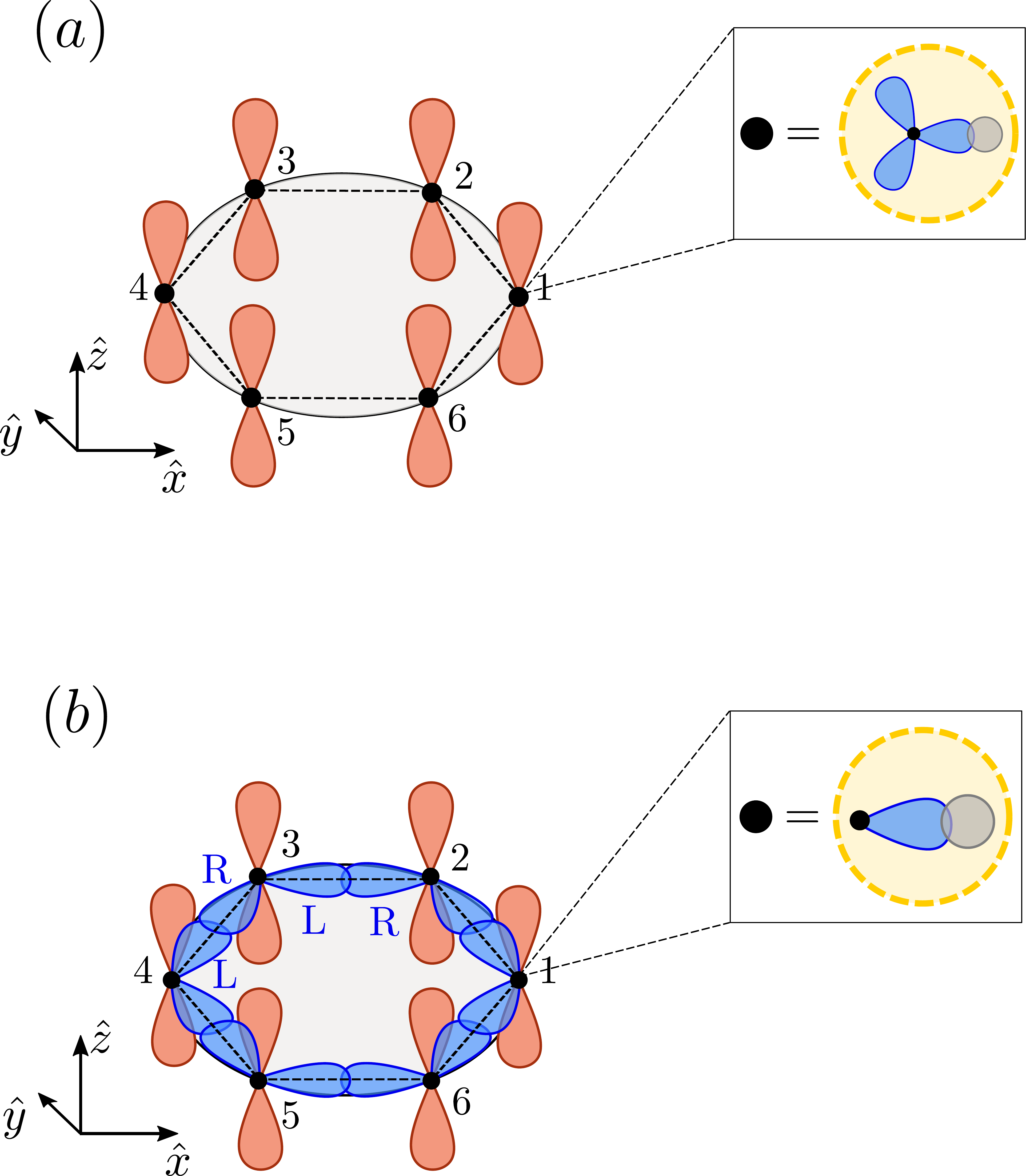}
\caption{Orbital structure of the rings we consider in this work. Here, we illustrate a ring with six sites, which consists of a prototype of the benzene molecule. (a) Single-band model considered in Sec.\ref{single}, with one $p_z$ orbital per site. In this case, only the degrees of freedom of the $\pi$-electrons are taken into account. The bonding $\sigma$-electrons are frozen and incorporated into the ring's sites, as illustrated in the inset, and they only contribute generate the static crystal potential. (b) Three-band model, with one $p_z$ orbital and two $sp^2$ orbitals, which we denominate \textit{left} $sp^2$ orbital ($L$) and \textit{right} $sp^2$ orbital ($R$) according to the right-hand rule. As shown in the inset of panel (b), the degrees of freedom of the third $sp^2$ orbital of each site, as well as those of the valence orbitals of another atom that might bind to it (hydrogen atom, in the case of a benzene molecule), are frozen and incorporated to the ring's sites. The ring's sites are always enumerated in ascending order in the counter-clockwise direction.}
\label{fig:1}
\end{figure}

Although we can try to make our arguments the most general possible, it is more instructive to work with a specific case where our ideas can be more clearly stated. Therefore, we consider our ring as a closed chain of atoms that are held together
by electrons in  \textit{covalent bonds}, as illustrated in Fig. \ref{fig:1}. For example, if we think of each site as a carbon atom, we can interpret them as prototypes of real aromatic molecules such as benzene if one hydrogen atom is attached to each carbon atom - see insets of Fig.\ref{fig:1}(a) and Fig.\ref{fig:1}(b) - or the recently synthesized cyclo[18]carbon \cite{c18}. In doing so, we are automatically pointing to a family of systems where our results can be tested. However, contrary to real-life molecules, we impose the sites to be always static. That is because, in this work, we focus solely on the orbital electronic properties and, thus, we do not investigate effects related to the ionic degrees of freedom, such as the molecular vibrational levels. Another remark is that our strategy can be equally applied to infinite chains of carbon atoms, which, if properly modified by the inclusion of hydrogen atoms, mimics, for example, the behavior of conducting polymers \cite{heeger}. Let us start by briefly reviewing some properties of the carbon orbitals.

A neutral carbon atom has a total of six electrons, two of which are in the $1s$ shell, strongly bound to the nucleus, while the remaining four electrons are in the outermost $2s$ and $2p$ orbitals \cite{cohen}. In the ring configuration (benzene molecule, in particular), the $2s$, $2p_{x}$ and $2p_y$ states of each carbon atom hybridize, defining three orthonormal $sp^{2}$ orbitals, whereas the $p_z$ orbital remains unchanged \cite{Baym}. The $sp^{2}$ orbitals are oriented along with directions in the ring's plane that make an angle of $2\pi/3$ between each other. The $p_z$ orbitals, on the other hand, are oriented perpendicularly to the ring's plane. The overlap between the $sp^2$ orbitals of two adjacent carbon atoms, as well as the overlap between a $sp^2$ orbital of a carbon atom and the $s$ orbital of a hydrogen atom in the specific case of the benzene molecule, form \textit{covalent bonds} known as $\sigma$-bonds. Moreover, the overlap between neighboring $p_z$ orbitals form the so-called  $\pi$-bonds, a weaker type of covalent bond. Briefly speaking, the $\pi$-bonds are weaker than the $\sigma$-bonds because the overlap between adjacent $p_z$ orbitals are much smaller than that of neighboring $sp^2$ orbitals \footnote{There is no overlap between $p_z$ and $sp^2$ orbitals for planar rings since they have opposite parity.}. Following the usual nomenclature, we hereafter denominate the electrons at the $sp^2$ orbitals by \textit{$\sigma$-electrons}, while those occupying the $p_z$ orbitals are called \textit{$\pi$-electrons}.

Hereafter, we focus on microscopic rings with a small number of sites, $3\leq N\leq 6$, mostly because, in these cases, we are able to perform exact diagonalization of the Hamiltonians we study in the subsequent sections. However, the extended Hubbard model we derive in Sec.\ref{H0new}, which is one of the central results of this paper, holds for any number of sites $N$ and can also be extended to $2D$ carbon lattices, such as graphene. Such a generalization, however, is left for future work, since we are here mainly interested in $1D$ systems.

We choose the ring's plane coinciding with the $xy$ plane, adopt its center as the origin of the coordinate system and align the site 1 with the $x$ axis. In this configuration, the position of the $j$-th site of the ring is given by 
\begin{equation}
\mbox{\boldmath$\mathcal{R}$}_{j}=\frac{a}{\sqrt{2\left(1-\cos(2\pi/N)\right)}}\left[\cos\alpha_j\hat{\bf{x}}+\sin\alpha_j\hat{\bf{y}}\right] \text{ ,}
\label{eq:R1}
\end{equation}

\noindent with $j=1,2,\cdots N$. Here, $\alpha_j=2(j-1)\pi/N$ denotes the angular position of site $j$, and $a$ is the lattice spacing. In analogy with the real aromatic molecules, we consider \textit{three orbitals per site}: one $p_z$ orbital, and two $sp^2$ orbitals. The third $sp^2$ orbital at each site, which binds it to another atom (such as the hydrogen atom in the case of the benzene molecule), is considered frozen, and therefore incorporated to the ring's sites, as illustrated in the insets of Fig.\ref{fig:1}(a) and \ref{fig:1}(b). The geometric bond configuration suggests that, whatever effect there might be of the $\sigma$-electrons on the dynamics of the $\pi$-electrons, it is much more likely that the $\sigma$-electrons contribute only with an effective local potential to the $\pi$-electrons. 

\subsection{Single-band Hubbard model}\label{single}
  
\begin{figure*}[t!]
\centering
\includegraphics[width=1\linewidth]{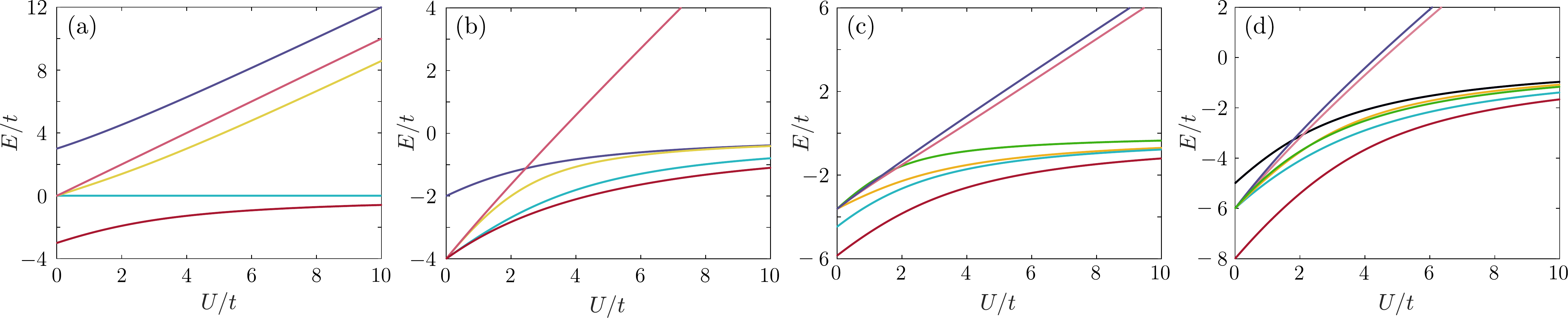}
\caption{Energy spectrum of the Hubbard Hamiltonian (Eq.(\ref{eq:H0})). The panels show the energy levels, as function of $U/t$ for rings with (a) $N=3$ sites, (b) $N=4$ sites, (c) $N=5$ sites and (d) $N=6$ sites at the half-filling regime ($N=N_e$). In panels (b) to (d), only a few of the low-lying energy levels are shown.}
\label{fig:levels0}
\end{figure*}  

Let us take a step back and start by describing only the degrees of freedom of the itinerant $\pi$-electrons. The simplest model for $N_{e}$ itinerant and \textit{interacting} electrons in a single orbital $N$-site lattice is the standard single-band Hubbard model\cite{Hubbard1}, 
\begin{equation}
\hat{H}_0=-t\sum\limits_{j=1}^{N}\sum\limits_{\sigma}\left(c_{j\sigma}^{\dag}c_{j+1\sigma}^{\null}+\text{ h.c.}\right)+U\sum\limits_{j=1}^{N}\hat{n}_{j\uparrow}\hat{n}_{j\downarrow} \text{ ,}
\label{eq:H0}
\end{equation}   

\noindent  where the operator $c_{j\sigma}^{\dag}$ ($c_{j,\sigma}$) creates (annihilates) an electron with spin $\sigma$ at the $p_z$ orbital of the $j$-th site of the ring, and $\hat{n}_{j\sigma}=c_{j\sigma}^{\dag}c_{j\sigma}^{\null}$ is the number operator. The parameter $t$ is the first-neighbor hopping \cite{Hubbard1}, which is given by
\begin{equation}
t=-\int d\mathbf{r}\;\varphi_{i+1}^{*}(\mathbf{r})\,\left[-\frac{\hbar^2}{2m}\boldsymbol{\nabla}^{2}+V_{c}(\mathbf{r})\right]\,\varphi_{i}(\mathbf{r}) \text{ ,}
\label{eq:t}
\end{equation}

\noindent Here $\varphi_{i}(\mathbf{r})$ is the Wannier wave function of an electron at the $p_z$ orbital of site $i$, and $V_{c}(\mathbf{r})$ is a periodic potential generated by \textit{the ions together with the core and bonding $\sigma$-electrons}(see Fig.\ref{fig:1}). In terms of these localized  wave functions, the on-site Coulomb repulsion takes the form
\begin{equation}
U=e^2\int d\mathbf{r}\int d\mathbf{r'}\;\frac{\left|\varphi_{i}(\mathbf{r})\right|^2\left|\varphi_{i}(\mathbf{r'})\right|^2}{\left|\mathbf{r}-\mathbf{r}\,'\right|} \text{ ,}
\label{eq:U} 
\end{equation}
  
\noindent where $e$ is the elementary electronic charge. If we set $U=0$, Eq.(\ref{eq:H0}) reduces to a purely tight-binding Hamiltonian, also  known as the \textit{H\"uckel Hamiltonian}\cite{Huckel1}. 

We do not specify the functional forms of $\varphi_j(\mathbf{r})$ and $V_{c}(\mathbf{r})$. The specific angular and radial dependence of $\varphi_j(\mathbf{r})$ is important to calculate the numerical values for the parameters $t$ and $U$. Here, it is enough to keep in mind that the deeper $V_{c}(\mathbf{r})$ is at the ionic positions, the larger is the tendency of the electrons to localize around those sites and, therefore, the smaller is the hopping amplitude. Besides, except when explicitly mentioned otherwise, all physical quantities calculated in this work are given in units of $t$. An estimate of hopping and on-site repulsion parameters for the specific case of the prototype of benzene is provided in Ref. \cite{thesis}. 

\subsection{Energy spectrum}\label{spectrum}

In the case of the H\"uckel Hamiltonian (Eq.(\ref{eq:H0}) with $U=0$), we can readily determine the energy levels and corresponding eigenstates of a generic ring with $N$ sites and $N_{e}$ \textit{independent} electrons. The transformation from \textit{site basis} to \textit{Bloch basis}, 
\begin{align}
&c_{j\sigma}^{\dag}=\frac{1}{\sqrt{N}}\sum\limits_{j=1}^{N}e^{-i2\pi kj/N}c_{k\sigma}^{\dag}\text{ ,}\\[0.2cm]
&c_{j\sigma}^{\null}=\frac{1}{\sqrt{N}}\sum\limits_{j=1}^{N}e^{i2\pi kj/N}c_{k\sigma}^{\null} \text{ ,}
\end{align}

\noindent where $c_{k\sigma}^{\dag}$ ($c_{k\sigma}^{\null}$) creates (annihilates) an electron with spin $\sigma$ and quasi-momentum $k$ (with $k=0,1,\cdots N-1$), diagonalizes the H\"uckel Hamiltonian:
\begin{equation}
\hat{H}_{\text{H\"uckel}}=-2t\sum\limits_{k=1}^{N-1}\sum\limits_{\sigma=\uparrow,\downarrow}\cos\left(\frac{2\pi k}{N}\right) c_{k\sigma}^{\dag}c_{k\sigma}^{\null}\text{ .}
\label{eq:Huckel_Bloch}
\end{equation}

\noindent Therefore, to build the many-body energy eigenstates we only need to fill up the \textit{single-particle} levels obeying the Pauli exclusion principle. Unfortunately, such a simple picture does not hold for $U\neq 0$, and numerical diagonalization is required. This procedure is not a trivial task, since the dimension $d = 2N!/[(2N-N_e)!N_e!]$ of the Fock space where $\hat{H}_0$ is defined grows faster than exponentially with the factorial of the number of sites and electrons of the ring.

Fig. \ref{fig:levels0} shows some of the energy levels, as a function of $U/t$, obtained through numerical diagonalization of Eq.(\ref{eq:H0}) for rings with (a) $N=3$, (b) $N=4$, (c) $N=5$ and (d) $N=6$ sites. These energy spectra were calculated for the rings in the \textit{half-filling regime}, where $N=N_e$. This choice is motivated by the electronic configuration of the benzene molecule, where we have a total of six $\pi$-electrons occupying the six $p_z$ orbitals of the aromatic ring.  Panels (b)-(d) show just a few of the low-lying energy levels of the systems. That is because their complete energy spectrum contains a large number of levels, and including all of their representative curves in the same panel results in quite cumbersome figures.

\section{Extended Hubbard model}\label{H0new}

The scenario we explored in  Sec.\ref{Model} suggests a separation of energy scales in the system, as follows. Since the $\sigma$-electrons are localized at the bonds, they can be described approximately by another Hubbard model, with a different hopping ($\tilde{t}$) between the left (L) and right (R) $sp^{2}$ orbitals of nearest neighbor sites (see Fig.\ref{fig:1}(b)) and a different on-site Hubbard interaction ($\tilde U$). Since these two orbitals are directed towards one another, we expect $\tilde{t}$ to be larger than the hopping $t$ of the $\pi$-electrons. Therefore, in each bond we have a two-electron - two-site Hubbard model with a larger hopping describing the interacting bonding electrons. In this picture, we consider bonding electrons at different bonds as distinguishable. The two lowest-lying energy eigenstates of the bond will be linear combinations of the Slater determinants of the symmetric and anti-symmetric linear combinations of the two above-mentioned neighboring orbitals.  By this reasoning,  it would be more costly to excite the bond to its first excited state in comparison to the characteristic energy of the $\pi$-electrons. In other words, the energy scale separating the ground state and the first excited state of the $\sigma$-electrons (let's denote it by $\Lambda$) is expected to be larger than the bandwidth of the $\pi$-electrons, which is set by the hopping parameter $t$ defined in Eq.(\ref{eq:t}). Therefore, recalling the uncertainty principle $\Delta E\Delta t\geq \hbar/2$, this implies that the $\sigma$-electrons excitations happen in a much faster time scale than that associated with the motion of the $\pi$-electrons around the ring. 

We can thus think of two different \textit{types} of electrons in the ring: the $\sigma$-electrons, which  are the \textit{fast} electrons, and  the $\pi$-electrons which are the \textit{slow} ones. This scenario resembles the well-known \textit{Born-Oppenheimer approximation} [see, for instance, Ref.\cite{Baym}] to decouple the nuclear and the electronic degrees of freedom of a molecule. Briefly speaking, due to the huge mass difference between the atomic nuclei and the electrons, the latter move around fixed positions of the former, which, when allowed to move, does it in a time scale much slower than that of the electronic motion. 

The application of the same arguments for the itinerant and bonding electrons seem to be very counter-intuitive because the latter are the ones to be localized at the chemical bonds, whereas the former are delocalized along the ring. Nevertheless, one should bear in mind that what must be really taken into account are the energy scales involved in the dynamics of each type of electron, and, indeed, the $\sigma$-electrons involve higher energy than the $\pi$-electrons. 

Here, guided by this energy scale separation, we use a perturbation approach, which we call \textit{generalized Born-Oppenheimer approximation}, in the sense that in our case, the degrees of freedom of the $\sigma$-electrons and the $\pi$-electron are those to be decoupled. It is fundamental to note that, in our approximation, contrary to the standard Born-Oppenheimer approximation, the ring's sites remain static all the time. No ionic degrees of freedom are addressed in our calculations. What we are aiming at is the effect of the dynamical distortion of the periodic potential felt by the $\pi$-electrons due to the excitation of $\sigma$-electrons only.
     
\subsection{Generalized Born-Oppenheimer approximation}\label{BO_Approx}

Here, it is more convenient to return to first quantization where the \textit{complete} Hamiltonian of a ring with $N$ sites and $N_{e}$ electrons, among which $N_{e}^{(\pi)}$ are $\pi$-electrons and $N_{e}^{(\sigma)}$ are $\sigma$ electrons, is given by $\mathcal{H}=\mathcal{H}_{p}+\mathcal{H}_{b}$. The term 
\begin{equation}
\mathcal{H}_{p}=\sum\limits_{i=1}^{N_{e}^{(\pi)}}\left(\frac{\mathbf{P}_{i}^{2}}{2m}+\tilde{V}_{c}(\mathbf{R}_{i})\right)+\frac{1}{2}\sum\limits_{i\neq j}U\left(\mathbf{R}_{i}-\mathbf{R}_{i}\right)
\label{eq:Hp}
\end{equation}

\noindent describes the $\pi$-electrons, with momenta and positions denoted by $\mathbf{R}_{i}$ and $\mathbf{P}_{i}$, respectively ($i=1,2,\cdots N_{e}^{(\pi)}$). In this equation, $U(\mathbf{r},\mathbf{r}')=e^{2}/\left|\mathbf{r}-\mathbf{r}'\right|$ is the standard Coulomb repulsion. Moreover, the Hamiltonian
\begin{align}
\mathcal{H}_{b}=&\sum\limits_{\alpha=1}^{N_{e}^{(\sigma)}}\left(\frac{\mathbf{p}_{\alpha}^{2}}{2m}+\tilde{V}_{c}(\mathbf{r}_{\alpha})\right)+\frac{1}{2}\sum\limits_{\alpha\neq\beta}U\left(\mathbf{r}_{\alpha}-\mathbf{r}_{\beta}\right)\nonumber\\
&+\sum\limits_{i,\alpha}U\left(\mathbf{r}_{\alpha}-\mathbf{R}_{i}\right)
\label{eq:Hb} 
\end{align}
accounts for both the degrees of freedom of  the $\sigma$-electrons, with momenta and positions denoted by $\mathbf{r}_{\alpha}$ and $\mathbf{p}_{\alpha}$ , respectively ($\alpha=1,2,\cdots N_{e}^{(\sigma)}$), and their coupling with the $\pi$-electrons. Hereafter, we reserve Roman (Greek) characters as indices for quantities referring to $\pi$-electrons ($\sigma$-electrons). It is important to note that the periodic potential $\tilde{V}_{c}(\mathbf{r})$ that appears in Eqs.(\ref{eq:Hp}) and (\ref{eq:Hb}) is not the same as $V_{c}(\mathbf{r})$ defined in Eq.(\ref{eq:t}). While $V_{c}(\mathbf{r})$ is generated by both the ring's sites with its core electrons, and the \textit{frozen $\sigma$-electrons in the bonds}, $\tilde{V}_{c}(\mathbf{r})$, on the other hand, do not include any contribution from  the $\sigma$-electrons. In other words, recalling our discussion in the beginning of this section, $V_{c}(\mathbf{r})$ is essentially $\tilde{V}_{c}(\mathbf{r})$ dressed by the static charge density in the bonds generated by the $\sigma$-electrons in their many-body ground state.

In this section, we denote by $\psi(\mathbf{r},\mathbf{R})$ the total many-body wave function of our system, where $\mathbf{r}$ stands for the entire set of positions of the $\sigma$-electrons $\{\mathbf{r}_{1},\mathbf{r}_{2},\cdots, \mathbf{r}_{N_{e}^{(\sigma)}}\}$, and similarly $\mathbf{R}$ denotes the set of positions of the $\pi$-electrons, $\{\mathbf{R}_{1},\mathbf{R}_{2},\cdots, \mathbf{R}_{N_{e}^{(\pi)}}\}$. Motivated by the aforementioned separation of energy scale, we assume that the total wave function has the following separable form:
\begin{equation}
\psi(\mathbf{r},\mathbf{R})=\sum\limits_{\nu}\phi_{\nu}(\mathbf{R})\,\varphi_{\nu}(\mathbf{r},\mathbf{R}) \text{ ,}
\label{eq:ansatz}
\end{equation}

\noindent where $\phi_{\nu}(\mathbf{R})$ refers to the $\pi$-electrons wave functions, and $\varphi_{\nu}(\mathbf{r},\mathbf{R})$ denotes the $\sigma$-electrons wave functions for a \textit{frozen configuration of $\pi$-electrons} (fixed $\mathbf{R}$). The latter obeys the following Schr\"odinger equation:
\begin{equation}
\mathcal{H}_{b}(\mathbf{R})\varphi_{\nu}(\mathbf{r},\mathbf{R})=\lambda_{\nu}(\mathbf{R})\varphi_{\nu}(\mathbf{r},\mathbf{R})\text{ .}
\label{eq:sigma_Sch}
\end{equation}

\noindent We emphasize that $\mathbf{R}$ in Eq.(\ref{eq:sigma_Sch}) is an external parameter rather than a dynamical variable. For each $\mathbf{R}$, the Schr\"odinger equation (\ref{eq:sigma_Sch}) determines the $\sigma$-electrons eigenvalues $\lambda_{\nu}(\mathbf{R})$ (with quantum numbers $\nu=0,1,2\cdots$), which, as it will shortly become clear, act as extra external potentials for the $\pi$-electrons. Note that, in principle, $\nu$ actually refers to a set $\{n_\alpha\}$ where $n_\alpha=0,1,2,...$ and $\alpha=1,2,...,3N_{e}^{(\sigma)}$. However, as we shall organize the energy levels in ascending order, we may label $\{n_\alpha\}$ as a sequence of integers $\nu$, specifying any eventual degeneracy whenever necessary.  

Substituting the ansatz (\ref{eq:ansatz}) into the full time-independent Schr\"odinger equation $\mathcal{H}\psi=E\psi$ and using Eq.(\ref{eq:sigma_Sch}), we find that the $\pi$-electrons wave function must obey
\begin{widetext}
\begin{align}
&\sum\limits_{\nu}\Bigg\{\left[\mathcal{H}_{p}\phi_{\nu}(\mathbf{R})+\lambda_{\nu}(\mathbf{R})\phi_{\nu}(\mathbf{R})\right]\varphi_{\nu}(\mathbf{r},\mathbf{R})+\frac{1}{2m}\sum\limits_{j=1}^{N_{e}^{(\pi)}}\left[\mathbf{P}_{j}^{2}\,\varphi_{\nu}(\mathbf{r},\mathbf{R})+2\left(\mathbf{P}_{j}\,\varphi_{\nu}(\mathbf{r},\mathbf{R})\right)\cdot\mathbf{P}_{j}\right]\Bigg\}\phi_{\nu}(\mathbf{R})=\nonumber\\
&E\sum\limits_{\nu}\phi_{\nu}(\mathbf{R})\varphi_{\nu}(\mathbf{r},\mathbf{R})\text{.}
\label{eq:phi_equation}
\end{align}
\end{widetext}
\noindent where $\mathbf{P}_{j}\,\varphi_{\nu}(\mathbf{r},\mathbf{R})$ is merely $-i \hbar\boldsymbol{\nabla}_{j}\,\varphi_{\nu}(\mathbf{r},\mathbf{R})$, the gradient of  $\varphi_{\nu}(\mathbf{r},\mathbf{R})$ with respect to ${\mathbf{R}}_i$ considered as a parameter in $\varphi_{\nu}(\mathbf{r},\mathbf{R})$.
Now, multiplying Eq.(\ref{eq:phi_equation}) on the left by $\varphi_{\mu}^{*}(\mathbf{r},\mathbf{R})$, integrating over the $\sigma$-electron positions, and using the fact that $\varphi_{\mu}(\mathbf{r},\mathbf{R})$ defines an orthonormal basis, i.e.,
\begin{equation}
\left\langle\varphi_{\mu}\left.\right|\varphi_{\nu}\right\rangle_{\mathbf{r}}=\int d\mathbf{r}\,\varphi_{\mu}^{*}(\mathbf{r},\mathbf{R})\varphi_{\nu}(\mathbf{r},\mathbf{R})=\delta_{\mu,\nu} \text{ ,}
\end{equation} 

\noindent we readily rewrite Eq.(\ref{eq:phi_equation}) as the following set of coupled equations   
\begin{equation}
\left[\mathcal{H}_{p}+\lambda_{\nu}(\mathbf{R})\right]\phi_{\nu}(\mathbf{R})+\sum_{\mu}\mathcal{A}_{\nu\mu}\phi_{\mu}(\mathbf{R})=E\phi_{\nu}(\mathbf{R}) \text{ .}
\label{eq:pi_Schrodinger}
\end{equation} 

\noindent Note that, in constrast to Eq.(\ref{eq:sigma_Sch}), $\mathbf{R}$ is now a dynamical variable. Moreover, the operator $\mathcal{A}_{\nu\mu}$ is responsible for coupling the $\pi$-electron wave functions with different $\mu$ and $\nu$, and has the form
\begin{equation}
\mathcal{A}_{\nu\mu}=f_{\nu\mu}(\mathbf{R})+\sum\limits_{j=1}^{N_{e}^{(\pi)}}\mathbf{g}_{\nu\mu}^{(j)}(\mathbf{R})\cdot \mathbf{P}_{j} \text{ ,}
\label{eq:Anu}
\end{equation}  

\noindent with
\begin{align}
&f_{\nu\mu}(\mathbf{R})\equiv-\frac{\hbar^{2}}{2m}\sum\limits_{j=1}^{N_{e}^{(\pi)}}\left\langle\varphi_{\nu}\left|\right.\boldsymbol{\nabla}_{j}^{2}\,\varphi_{\mu}\right\rangle_{\mathbf{r}}=\nonumber\\
&=-\sum\limits_{j=1}^{N_e^{(\pi)}}\frac{\hbar^{2}}{2m}\int d\mathbf{r}\,\varphi_{\nu}^{*}(\mathbf{r},\mathbf{R})\boldsymbol{\nabla}_{j}^{2}\,\varphi_{\mu}(\mathbf{r},\mathbf{R}) \text{ ,}\label{eq:fmunu}
\end{align}
and
\begin{align}
&\mathbf{g}_{\nu\mu}^{(j)}(\mathbf{R})\equiv-\frac{i\hbar}{m}\left\langle\varphi_{\nu}\left|\right.\boldsymbol{\nabla}_{j}\,\varphi_{\mu}\right\rangle_{\mathbf{r}}\nonumber\\
&=-\frac{i\hbar}{m}\int d\mathbf{r}\,\varphi_{\nu}^{*}(\mathbf{r},\mathbf{R})\boldsymbol{\nabla}_{j}\,\varphi_{\mu}(\mathbf{r},\mathbf{R}) \label{eq:gmunu}\text{.}
\end{align}

\begin{figure}[b!]
\centering
\includegraphics[width=0.8\linewidth]{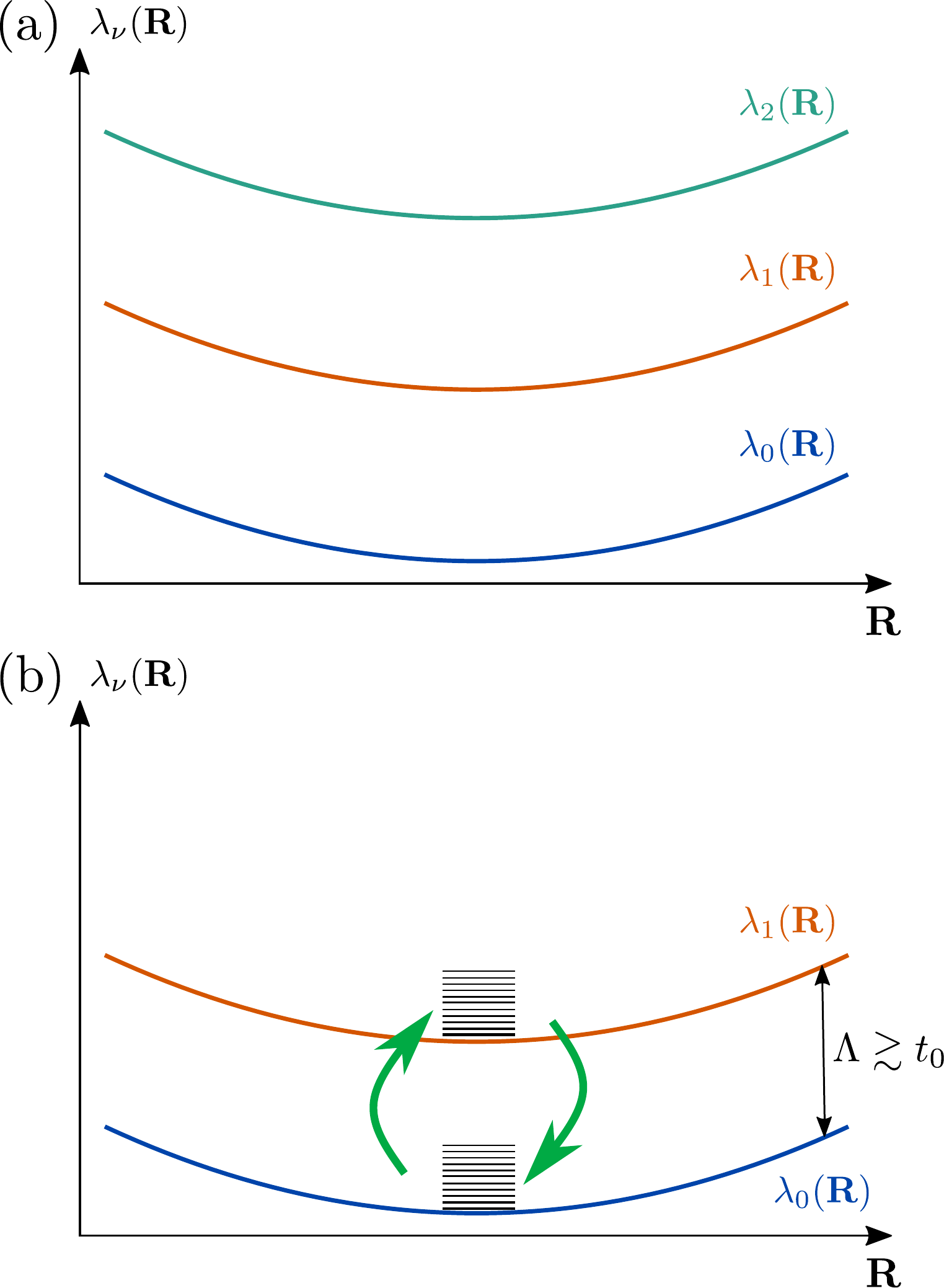}
\caption{Energy "surfaces" of the $\sigma$-electrons. Illustration of the energy levels of the $\sigma$-electrons as a function of the $\pi$-electrons configuration $\lambda_{\nu}(\mathbf{R})$ as if they were a function of a scalar variable, in analogy to the simpler standard Born-Oppenheimer approximation. Panel (a) represents the first three low-lying $\lambda_{\nu}(\mathbf{R})$. Panel (b) focuses only on the first two $\sigma$-electrons energy levels. In each one of them, the $\pi$-electron Hubbard spectrum is represented by the horizontal black lines. The blue arrows indicate virtual excitations that may occur in the system if the energy separation ($\Lambda$) between the two $\sigma$-electron surfaces is comparable with the $\pi$-electrons bandwidth, which is set by the $\pi$-electron hopping amplitude $t$.}
\label{fig:BO}
\end{figure}

In order to develop a more intuitive picture of the meaning of Eqs.(\ref{eq:sigma_Sch}) and (\ref{eq:pi_Schrodinger}) we show a simplified sketch of $\lambda_{\nu}(\mathbf{R})$ in Fig. \ref{fig:BO}. Panel (a) illustrates three of these eigenvalues as if they were a function of a scalar variable, in analogy with the standard Born-Oppenheimer approximation, to which we are more used. In reality, of course, $\lambda_{\nu}(\mathbf{R})$ defines a hypersurface in the space configuration of the $\pi$-electrons. Hereafter, since we want to derive a \textit{low-energy effective model for the $\pi$-electrons}, we focus only on the ground state and the first excited state of the $\sigma$-electrons, as illustrated in Fig. \ref{fig:BO}(b), i.e., we truncate the sum in Eq.(\ref{eq:ansatz}) at $\nu=1$. In this case, Eq.(\ref{eq:pi_Schrodinger}) simplifies to a set of two coupled equations, 
\begin{align}
&\mathcal{H}_{0}\,\phi_{0,n}(\mathbf{R})+\mathcal{A}_{01}\,\phi_{1,n}(\mathbf{R})=E_{n}\,\phi_{0,n}(\mathbf{R}) \text{ ,}\label{eq:A}\\[0.2cm]
&\mathcal{H}_{1}\,\phi_{1,n}(\mathbf{R})+\mathcal{A}_{10}\,\phi_{0,n}(\mathbf{R})=E_{n}\,\phi_{1,n}(\mathbf{R})\text{ .}\label{eq:B}
\end{align}

\noindent Here, index $n$ labels the $\pi$-electron quantum numbers. Besides, we define $\mathcal{H}_0\equiv \mathcal{H}_{p}+\lambda_0(\mathbf{R})+\mathcal{A}_{00}$ and $\mathcal{H}_1\equiv \mathcal{H}_{p}+\lambda_1(\mathbf{R})+\mathcal{A}_{11}$. We emphasize that in the language of second quantization, $\mathcal{H}_0$ is a single-band Hubbard Hamiltonian with a renormalized hopping amplitude $t_0$. Similarly, $\mathcal{H}_1$ is a Hubbard Hamiltonian with another hopping parameter $t_1$.

If $\lambda_{0}(\mathbf{R})$ and 
$\lambda_{1}(\mathbf{R})$ are energetically too far apart, ($\Lambda_{1,0}(\mathbf{R})\equiv \lambda_{1}(\mathbf{R})-\lambda_0(\mathbf{R})\gg t_0$),  $\mathcal{A}_{\nu\mu}$ becomes negligible. Consequently, Eqs. (\ref{eq:A}) and (\ref{eq:B}) decouple and the system's energy levels are just the set composed by the union of the eigenvalues of $\mathcal{H}_0$ and $\mathcal{H}_{1}$, illustrated by the horizontal black lines in Fig. \ref{fig:BO}(b). In this case, the low-lying energy states of the system are those of $\mathcal{H}_0$, which means that the $\pi$-electrons move along the ring as if the $\sigma$-electrons were frozen in their ground state $\lambda_0(\mathbf{R})$, recovering the standard Hubbard model we described in Sec.\ref{single}. The interesting limit we consider here is when $\Lambda_{1,0}(\mathbf{R})$ is still larger than $t_0$, but they are of the \textit{same order} ($\Lambda_{1,0}(\mathbf{R})\gtrsim t_0$). In this case, $\mathcal{A}_{\nu\mu}$ mixes the eigenstates of $\mathcal{H}_0$ and $\mathcal{H}_1$. Let's explore this scenario in more details in the next paragraphs.
      
Isolating $\phi_{1,n}(\mathbf{R})$ in Eq.(\ref{eq:B}) and substituting it in Eq.(\ref{eq:A}), we obtain an effective Schr\"odinger equation for $\phi_{0,n}(\mathbf{R})$,
\begin{equation}
\left[\mathcal{H}_{0}+\mathcal{A}_{01}\left(E_n-\mathcal{H}_{1}\right)^{-1}\mathcal{A}_{10}\right]\phi_{0,n}(\mathbf{R})=E_{n}\,\phi_{0,n}(\mathbf{R}) \text{ .}
\label{eq:effective0}
\end{equation} 

\noindent Here 
\begin{equation}
\mathcal{W}_{eff}(\mathbf{P},\mathbf{R})\equiv\mathcal{A}_{01}\left(E_n-\mathcal{H}_{1}\right)^{-1}\mathcal{A}_{10} \text{ ,}
\label{eq:potential_eff}
\end{equation}

\noindent which in general depends on both momenta and positions, defines an \textit{effective interaction} between the $\pi$-electrons, which carries information about the virtual excitations of the $\sigma$-electrons. Moreover, Eq.(\ref{eq:effective0}) is a self-consistent equation, since the potential defined in Eq.(\ref{eq:potential_eff}) itself depends on the energy levels $E_{n}$ we want to calculate. Fortunately, this is a typical problem that can be approached by the well-known \textit{Wigner-Brillouin perturbation theory} \cite{Baym}.

Let us denote by $\zeta_{0,n}(\mathbf{R})$ and $\varepsilon_{n}^{(0)}$ ($\zeta_{1,n}(\mathbf{R})$ and $\varepsilon_{n}^{(1)}$) the eigenstates and corresponding eigenvalues of the Hubbard-like Hamiltonian $\mathcal{H}_0$ ($\mathcal{H}_1$). Both $\zeta_{0,n}(\mathbf{R})$ and $\zeta_{1,n}(\mathbf{R})$ span an orthonormal basis, i.e.
\begin{align}
&\sum\limits_{n}\left|\zeta_{\nu,n}\right\rangle\left\langle \zeta_{\nu,n}\right|=\mathbb{1}\text{ ,}\label{eq:1}\\
&\left\langle \zeta_{\nu,n}\left|\zeta_{\nu,m}\right.\right\rangle=\int d\mathbf{R}\zeta_{\nu,n}^*(\mathbf{R})\zeta_{\nu,m}(\mathbf{R})=\delta_{n,m} \text{ ,}
\end{align}  

\noindent with $\nu=0,1$ and $\left\langle \zeta_{0,n}\left|\zeta_{1,m}\right.\right\rangle\neq \delta_{m,n}$. Wigner-Brillouin perturbation theory tells us that $\phi_{0,n}(\mathbf{R})$ and $\zeta_{0,n}(\mathbf{R})$, as well as $E_n$ and $\varepsilon_n$ are related through, up to second order in $\mathcal{W}_{eff}$,
\begin{align}
&\phi_{0,n}(\mathbf{R})=\zeta_{0,n}(\mathbf{R})+\sum\limits_{m\neq n}\frac{\left\langle \zeta_{0,m}\left|\mathcal{W}_{eff}\right|\zeta_{0,n}\right\rangle}{E_{n}-\varepsilon_{m}^{(0)}}\,\zeta_{0,m}(\mathbf{R})\\
&E_{n}=\varepsilon_n^{(0)}+ \left\langle \zeta_{0,n}\left|\mathcal{W}_{eff}\right|\zeta_{0,n}\right\rangle  \text{ .}
\label{eq:WBPT_energy}
\end{align} 

\noindent The matrix element of the effective interaction (\ref{eq:potential_eff}) is explicitly given by
\begin{equation}
\left\langle \zeta_{0,m}\left|\mathcal{W}_{eff}\right|\zeta_{0,n}\right\rangle=\int d\mathbf{R}\,\zeta_{0,m}^*(\mathbf{R})\mathcal{W}_{eff}(\mathbf{P},\mathbf{R})\,\zeta_{0,n}(\mathbf{R}) \text{ ,}
\label{eq:W_matrix}
\end{equation}

\noindent in the basis spanned by the $\zeta_{0,n}(\mathbf{R})$ states. 

In zeroth order perturbation theory for the energy ($E_{n}\approx\varepsilon_n^{(0)}$), and neglecting quadratic or higher orders of $\mathcal{W}_{eff}$ in the perturbation expression for the eigenstates, we obtain
\begin{align}
&\phi_{0,n}(\mathbf{R})\approx \zeta_{0,n}(\mathbf{R})
+\sum\limits_{m\neq n}\frac{1}{\varepsilon_{n}^{(0)}-\varepsilon_{m}^{(0)}}\nonumber\\
&\times\left\langle \zeta_{0,m}\left|\mathcal{A}_{01}\left(\varepsilon_{n}^{(0)}-\mathcal{H}_{1}\right)^{-1}\mathcal{A}_{10}\right|\zeta_{0,n}\right\rangle \zeta_{0,m}(\mathbf{R}) \text{ ,}
\label{eq:PT}
\end{align}

\noindent from which it is clear that the matrix element defined in Eq.(\ref{eq:W_matrix}) simplifies to 
\begin{equation}
\left\langle \zeta_{0,m}\left|\mathcal{W}_{eff}\right|\zeta_{0,n}\right\rangle\approx -\frac{1}{\Lambda}\left\langle \zeta_{0,m}\left|\mathcal{A}_{01}\mathcal{O}_{n}\mathcal{A}_{10} \right|\zeta_{0,n}\right\rangle \text{ .}
\label{eq:W_matrix_approx}
\end{equation}

\noindent We denote by $\mathcal{O}_{n}$ the many-body operator
\begin{equation}
\mathcal{O}_{n}\equiv \sum\limits_{m}\left(1-\frac{\varepsilon_{n}^{(0)}-\varepsilon_{m}^{(0)}}{\Lambda}\right)^{-1}\left|\zeta_{1,m}\right\rangle\left\langle\zeta_{1,m}\right| \text{ .}
\label{eq:O}
\end{equation}

\noindent To derive Eqs.(\ref{eq:W_matrix_approx}) and (\ref{eq:O}), we use the closure relation in Eq.(\ref{eq:1}) to rewrite $\varepsilon_{n}^{(0)}-\mathcal{H}_{1}$ in Eq.(\ref{eq:PT}) as 
\begin{align}
&\varepsilon_{n}^{(0)}\mathbb{1}-\sum\limits_{m}\varepsilon_{m}^{(1)}\left|\zeta_{1,m}\right\rangle\left\langle\zeta_{1,m}\right|=\nonumber\\
&\sum\limits_{m}\left(\varepsilon_{n}^{(0)}-\varepsilon_{m}^{(1)}\right)\left|\zeta_{1,m}\right\rangle\left\langle\zeta_{1,m}\right|\text{ .}
\end{align}   

\noindent Besides, we approximate the energy levels of $\mathcal{H}_{1}$ as those of $\mathcal{H}_0$ displaced by the energy separation between the two $\sigma$-electrons energy surfaces, i.e.
\begin{equation}
\varepsilon_{m}^{(1)}\approx \varepsilon_{m}^{(0)}+\Lambda_{1,0}(\mathbf{R}) \text{ .} 
\end{equation}

\noindent Recall that we previously define $\Lambda_{1,0}(\mathbf{R})\equiv \lambda_{1}(\mathbf{R})-\lambda_{0}(\mathbf{R})$. Interestingly, in \cite{thesis} it is shown that such energy spacing between the $\sigma$-electrons energy surface depends weakly on $\mathbf{R}$, so it is reasonable to approximate it by a constant, $\Lambda_{1,0}(\mathbf{R})\approx \Lambda>0$, consistently with the notation we have been using since the beginning of this section.  

Unfortunately, even after the aforementioned approximations, the effective interaction is still complicate because of the infinite sum involving the projectors in Eq.(\ref{eq:O}). Thus, to proceed further, we need to establish new simplifying hypotheses, and approximations, as described in detail in Appendix~\ref{first}, which, when applied to  Eqs.(\ref{eq:Anu})-(\ref{eq:gmunu}), lead us to
\begin{align}
&\mathcal{A}_{01}\approx-\frac{2i\hbar U g_{N}}{ma\Lambda}\sum\limits_{j=1}^{N_{e}^{(\sigma)}}\hat{\mathbf{n}}_{j}\cdot \mathbf{P}_{j} \text{ ,}\label{eq:a01}\\[0.2cm]
&\mathcal{A}_{10}\approx\frac{2i\hbar U g_{N}}{ma\Lambda}\sum\limits_{j=1}^{N_{e}^{(\sigma)}}\hat{\mathbf{n}}_{j}\cdot \mathbf{P}_{j} \label{eq:a10}\text{ .}
\end{align} 
 
\noindent Here $\hat{\bf{n}}_{j}\equiv (\hat{\mathbf{d}}_{j}^{(R)}-\hat{\mathbf{d}}_{j-1}^{(R)})/|\hat{\mathbf{d}}_{j}^{(R)}-\hat{\mathbf{d}}_{j-1}^{(R)}|$, with $|\hat{\mathbf{d}}_{j}^{(R)}-\hat{\mathbf{d}}_{j-1}^{(R)}|=g_{N}=\sqrt{2+2\cos(2\pi/N)}$ is the versor in the direction of the position of the site at which the $\pi$-electron is localized, but pointing inwards.
 
At this point, we have almost everything we need to derive a simplified expression for $\mathcal{W}_{eff}$. All we need now is return  to Eq.(\ref{eq:O}), and analyze it more carefully. If $\mathcal{O}_{n}$ were a constant, it would generate a $\mathcal{W}_{eff}$ which would be just a product of two one-body operators, instead of a true two-body operator. However, the very form of $\mathcal{O}_{n}$, if written in coordinate representation, induces us to assume it is a non-separable function of the generalized coordinates $\bf{R}$ and $\bf{R'}$. Therefore,  the simplest assumption we can make about Eq.(\ref{eq:O}) is to neglect contributions from three, four-body interactions and so on, that is, it only has a two-body component which can correlate the momentum operators that appear in Eqs.(\ref{eq:a01}) and (\ref{eq:a10}). In this case, we obtain
\begin{equation}
\mathcal{W}_{eff}\approx-\frac{1}{\Lambda^3}\left(\frac{2\hbar U g_{N}}{ma}\right)^2\,\sum\limits_{i,j=1}^{N_{e}}\mathbf{P}_{i}\cdot\hat{\mathbf{n}}_{i}\mathcal{O}(\mathbf{R}_{i},\mathbf{R}_{j})\hat{\mathbf{n}}_{j}\cdot \mathbf{P}_{j} \text{ .}
\label{eq:Weff_final}
\end{equation}

\noindent  Note that since $\hat{\bf{n}}_{j}$ is a simple versor rather than an operator, we can freely interchange it with the momentum operator, i.e. $\hat{\bf{n}}_{j}\cdot \mathbf{P}_{j}=\mathbf{P}_{j}\cdot\hat{\bf{n}}_{j}$, and define a tensor 
\begin{equation}
\overleftrightarrow{T}(\mathbf{R}_{i},\mathbf{R}_j)\equiv \hat{\mathbf{n}}_{i}\mathcal{O}(\mathbf{R}_{i},\mathbf{R}_{j})\hat{\mathbf{n}}_{j} \text{ ,}
\label{eq:T}
\end{equation}

\noindent which encodes the information about the ring's $\sigma$-bonds orientation through the versors $\hat{\bf{n}}_{i}$. 

An effective attractive momentum-momentum interaction with a form similar to that of Eq.(\ref{eq:Weff_final})  appeared in the literature some decades ago, when Bohm and Pines wrote the seminal series of papers about the electron gas \cite{PinesI,PinesII,PinesIII}. They were able to show that there is an effective inter-electronic potential mediated by plasmons (longitudinal plasma fluctuations), which they called a \textit{residual interaction} \cite{PinesIII}. However, they argued that such an interaction is negligible because of the screening effects in a dense electron gas. In our case, on the other hand, since we are dealing with a few body systems, screening effects are not strong enough to suppress this kind of interaction. 

Only for the sake of completeness, it is worth mentioning that in Ref. \cite{PinesI} the authors also described the effective inter-electronic interaction mediated by transverse electromagnetic radiation in the electronic medium. This also results in a momentum-momentum interaction, that they recognized as a Biot-Savart interaction. The latter is basically unscreened and is a relativistic correction of the order of $(v/c)^2$ to the inter-electronic interaction. Incidentally, in classical electrodynamics, this same interaction is also described by the so-called  Breit-Darwin (or current-current) interaction \cite{jackson}. 
 
\subsection{Effective interaction in second quantization}\label{W2Q}

In the previous section, we show that virtual excitations of the $\sigma$-electrons mediate an effective momentum-momentum attraction between the $\pi$-electrons, which, in first quantization, is given by Eq.(\ref{eq:Weff_final}). Here, we derive its expression in the language of second quantization. By adding the second-quantized $W_{eff}$ to Eq.(\ref{eq:H0}), we derive an extended Hubbard Hamiltonian for the degrees of freedom of the $\pi$-electrons alone, but which takes into account the effects of the $\sigma$-electrons in their dynamics. It is important to note that in this section $\mathbf{r}$ no longer denotes the set of positions of the $\sigma$-electrons but rather a generic position in space.

Since $W_{eff}$ in Eq.(\ref{eq:Weff_final}) is a two-body operator, the standard procedure to determine its second-quantized expression is \cite{Fetter,Baym} 
\begin{widetext}
\begin{equation}
\hat{W}_{eff}=\frac{1}{2}\sum\limits_{\sigma,\sigma'}\int\int d\mathbf{r}\,d\mathbf{r'}\,\hat{\psi}_{\sigma}^{\dag}(\mathbf{r})\,\hat{\psi}_{\sigma'}^{\dag}(\mathbf{r'})\mathbf{P}\cdot \overleftrightarrow{T}(\mathbf{r},\mathbf{r'})\cdot\mathbf{P'}\,\hat{\psi}_{\sigma'}^{\null}(\mathbf{r'})\hat{\psi}_{\sigma}^{\null}(\mathbf{r}) \text{ ,}
\label{eq:W2Q}
\end{equation}
\end{widetext}

\noindent where,  in coordinate representation,  $\mathbf{P}=-i\hbar\boldsymbol{\nabla}$ and
 $\bf{P'}=-i\hbar\boldsymbol{\nabla'}$, with $\boldsymbol{\nabla'}$ denoting the gradient with respect to $\mathbf{r'}$. Besides, $\hat{\psi}_{\sigma}^{\dag}(\mathbf{r})$ ($\hat{\psi}_{\sigma}^{\null}(\mathbf{r})$) is the field operator that creates (annihilates) an electron with spin $\sigma$ at the position $\mathbf{r}$. Since we are deriving an effective model for the $\pi$-electrons alone, such a field operator is defined only in terms of the Wannier wave functions of the $p_z$ orbitals ($\varphi_{j\sigma}(\mathbf{r})$), as $\hat{\psi}_{\sigma}(\mathbf{r})=\sum\limits_{i,\sigma}^{}\varphi_{i\sigma}(\mathbf{r}) c_{i\sigma}$, where $c_{i\sigma}$ annihilates an electron with spin $\sigma$ at the site $i$. Here the reader should be warned not to confuse $\varphi_{j}(\mathbf{r})$ with the $\sigma$-electrons wave functions $\varphi_{\nu}(\mathbf{r},\mathbf{R})$ we defined in Sec.\ref{BO_Approx}.

Substituting the expression for $\hat{\psi}_{\sigma}(\mathbf{r})$ in terms of $c_{i\sigma}$ into Eq.(\ref{eq:W2Q}) we find the second-quantized effetive interaction in the \textit{site basis},
\begin{equation}
\hat{\mathcal{W}}_{eff}=-\frac{1}{2\Lambda^3}\left(\frac{2\hbar Ug_{N}}{ma}\right)^2\sum\limits_{i,j,k,l=1}^{N}\;\sum\limits_{\sigma,\sigma'}w_{ijkl}\,c_{i\sigma}^{\dag}c_{j\sigma'}^{\dag}c_{k\sigma'}^{\null}c_{l\sigma}^{\null} \text{ ,}
\label{eq:W_eff_2Q_site}
\end{equation}

\noindent where $w_{ijkl}$ is the matrix element
\begin{widetext}
\begin{equation}
w_{ijkl}\equiv \left\langle i j\left|\mathbf{P}\cdot \overleftrightarrow{T}\cdot \mathbf{P'}\right|l k\right\rangle=-\hbar^2\int\int d\mathbf{r}d\mathbf{r'}\,\varphi_{i}^{*}(\mathbf{r})\,\varphi_{j}^{*}(\mathbf{r'})\boldsymbol{\nabla}\cdot \overleftrightarrow{T}(\mathbf{r},\mathbf{r'}) \cdot\boldsymbol{\nabla'}\varphi_{k}^{\null}(\mathbf{r'})\varphi_{l}^{\null}(\mathbf{r})  \text{ ,}
\end{equation}
\end{widetext}
which we study in detail henceforth. 

Applying the closure identity, $\sum\limits_{i,j=1}^{N}\left|ij\right\rangle\left\langle ij\right|=\mathbb{1}$, between the momentum operators and the tensor $\overleftrightarrow{T}$, we obtain
\begin{equation}
w_{ijkl}=\sum\limits_{i_1,i_2}\sum\limits_{j_1,j_2}\left\langle ij\left|\mathbf{P}\right|i_1i_2\right\rangle\cdot\left\langle i_1i_2 \left|\overleftrightarrow{T}\right|j_1j_2\right\rangle\cdot\left\langle j_1j_2\left|\mathbf{P'}\right|lk\right\rangle\!.
\label{eq:w1}
\end{equation}

\noindent Note that $\mathbf{P}$ acts only on the \textit{first} entry of a ket $\left|ij\right\rangle$, i.e.
\begin{equation}
\left\langle \mathbf{r}\mathbf{r'}\left|\mathbf{P}\right|ij\right\rangle=\left\langle \mathbf{r}\left|\mathbf{P}\right|i\right\rangle\left\langle\mathbf{r'}\left|\right. j\right\rangle \text{ .}
\end{equation}

\noindent Similarly, $\mathbf{P'}$ acts only on the \text{second} entry of $\left|ij\right\rangle$. Therefore
\begin{align}
&\left\langle ij\left|\mathbf{P}\right|i_1i_2\right\rangle=\left\langle i\left|\mathbf{P}\right|i_1\right\rangle\left\langle j\left|\right. i_2\right\rangle=\left\langle i\left|\mathbf{P}\right|i_1\right\rangle\delta_{j,i_2} \text{ ,}\\[0.2cm]
&\left\langle j_1j_2\left|\mathbf{P'}\right|lk\right\rangle=\left\langle j_2\left|\mathbf{P'}\right|k\right\rangle \left\langle j_1\left|\right. l\right\rangle=\left\langle j_2\left|\mathbf{P'}\right|k\right\rangle\delta_{j_{1},l}  \text{ ,}
\end{align}

\noindent and, as a consequence of the orthonormality of the Wannier wave functions, Eq.(\ref{eq:w1}) becomes
\begin{equation}
w_{ijkl}=\sum\limits_{i_1,j_2=1}^{N}\sum\left\langle i\left|\mathbf{P}\right|i_1\right\rangle\cdot\left\langle i_1\,j \left|\overleftrightarrow{T}\right|l\,j_2\right\rangle\cdot\left\langle j_2\left|\mathbf{P'}\right|k\right\rangle \text{ .}
\label{eq:w2}
\end{equation}   

Consistently with the standard approximations used to derive the single-band Hubbard Hamiltonian \cite{Hubbard1}, the momentum matrices elements in Eq.(\ref{eq:w2}) can be approximated by a term connecting only nearest neighbor sites, 
\begin{equation}
    \left\langle i  \left|\mathbf{P}\right|j\right\rangle\approx \frac{imt}{\hbar}\left(\mbox{\boldmath$\mathcal{R}$} _{i}-\mbox{\boldmath$\mathcal{R}$} _{j}\right)\delta_{j,i\pm 1}\text{ .}
\label{eq:Pij}
\end{equation}

\noindent A detailed derivation of Eq.(\ref{eq:Pij}) is shown in Appendix \ref{P}.
Recall that $t$ is the hopping parameter between two neighboring $p_z$ orbital, and $m$ is the electron mass. Therefore, substituting Eq.(\ref{eq:Pij}) into Eq.(\ref{eq:w2}) we obtain four contributions for $w_{ijkl}$:
\begin{widetext}
\begin{eqnarray}
    &&w_{ijkl}\approx -\left(\frac{mt}{\hbar}\right)^2\times\nonumber\\
    &&\times \left[\left(\mbox{\boldmath$\mathcal{R}$} _{i}-\mbox{\boldmath$\mathcal{R}$} _{i+1}\right)\cdot\left\langle i+1\;\;j\left|\overleftrightarrow{T}\right|l\;\;k+1\right\rangle\cdot\left(\mbox{\boldmath$\mathcal{R}$} _{k+1}-\mbox{\boldmath$\mathcal{R}$} _{k}\right)+\left(\mbox{\boldmath$\mathcal{R}$} _{i}-\mbox{\boldmath$\mathcal{R}$} _{i+1}\right)\cdot\left\langle i+1\;\;j\left|\overleftrightarrow{T}\right|l\;\;k-1\right\rangle\cdot\left(\mbox{\boldmath$\mathcal{R}$} _{k-1}-\mbox{\boldmath$\mathcal{R}$} _{k}\right)\right.\nonumber\\
    &&\left.+\left(\mbox{\boldmath$\mathcal{R}$} _{i}-\mbox{\boldmath$\mathcal{R}$} _{i-1}\right)\cdot\left\langle i-1\;\;j\left|\overleftrightarrow{T}\right|l\;\;k+1\right\rangle\cdot\left(\mbox{\boldmath$\mathcal{R}$} _{k+1}-\mbox{\boldmath$\mathcal{R}$} _{k}\right)+\left(\mbox{\boldmath$\mathcal{R}$} _{i}-\mbox{\boldmath$\mathcal{R}$} _{i-1}\right)\cdot\left\langle i-1\;\;j\left|\overleftrightarrow{T}\right|l\;\;k-1\right\rangle\cdot\left(\mbox{\boldmath$\mathcal{R}$} _{k-1}-\mbox{\boldmath$\mathcal{R}$} _{k}\right)\right]\text{ .}\nonumber\\
    \label{eq:w3}
\end{eqnarray}

Assuming that the on-site contribution of the matrix elements of the tensor $\overleftrightarrow{T}$ is the dominant one, Eq.(\ref{eq:w3}) becomes (see Appendix \ref{w}) 
\begin{equation}
w_{ijkl}\approx\left(\frac{mt}{\hbar}\right)^2\frac{a^2\mathcal{O}_0 \left(1-\cos(2\pi/N)\right)}{2}\left[\delta_{j,i+1}\delta_{l,i+1}\delta_{k,i}+\delta_{j,i+1}\delta_{l,i+1}\delta_{k,i+2}
+\delta_{j,i-1}\delta_{l,i-1}\delta_{k,i-2}+\delta_{j,i-1}\delta_{l,i-1}\delta_{k,i}\right] \text{ ,}
\label{eq:w5}
\end{equation}  

\noindent leading to the following second-quantized expression for the effective interaction between the $\pi$-electrons:
\begin{equation}
\hat{W}_{eff}=-\frac{\left(tU\right)^2}{\Lambda^{3}}2\mathcal{O}_0\left(1-\cos^2(2\pi/N)\right)\sum\limits_{j=1}^{N}\sum\limits_{\sigma,\sigma'}\left[\left(c_{j\sigma}^{\dag}c_{j+1\sigma'}^{\dag}c_{j\sigma'}^{\null}c_{j+1\sigma}^{\null}+\text{h.c.}\right)
+\left(c_{j\sigma}^{\dag}c_{j-1\sigma'}^{\dag}c_{j-2\sigma'}^{\null}c_{j-1\sigma}^{\null}+\text{h.c.}\right)\right] \text{ .}
\label{eq:Weff2}
\end{equation}
\end{widetext}

\noindent Here, we define the site-independent constant $\mathcal{O}_{0}\equiv \left\langle ii \left|\mathcal{O}(\mathbf{r},\mathbf{r}')\right| ii\right\rangle$. See Appendix \ref{w} for more details.

\begin{figure}[t!]
\centering
\includegraphics[width=1\linewidth]{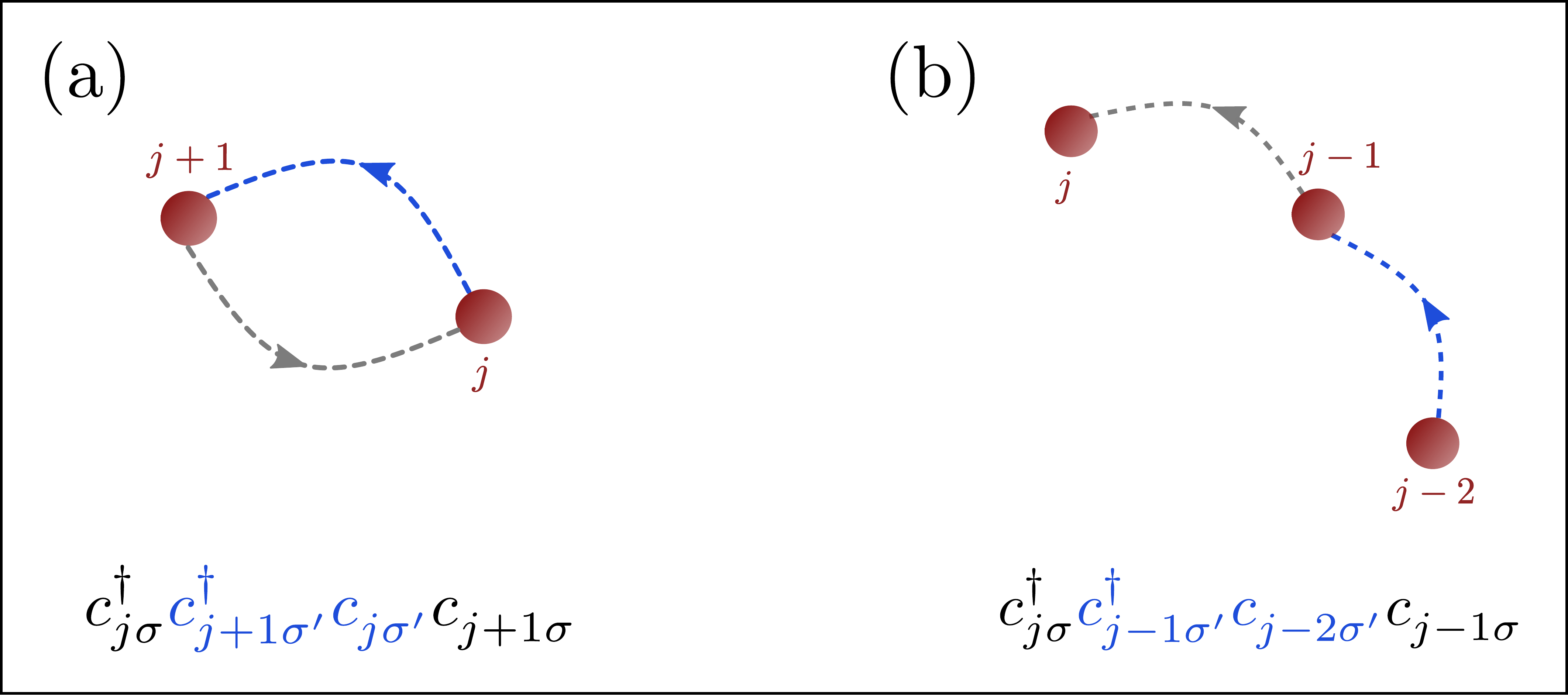}
\caption{\textbf{Effective interaction.} Illustration of two types of two-body processes that appear in the effective interaction Eq.(\ref{eq:Weff2}). (a) is the "Bubble term", while (b) is the extended term that favors the electron delocalization. The Hermitian conjugates of (a) and (b) just reverse the direction of the arrows.} 
\label{fig:processes}
\end{figure}

The effective interaction (\ref{eq:Weff2}) has two types of processes, as illustrated in Fig. \ref{fig:processes}. The first process, involving $ c_{j\sigma}^{\dag}c_{j+1\sigma'}^{\dag}c_{j\sigma'}^{\null}c_{j+1\sigma}^{\null} $ and its Hermitian conjugate, is what we called "bubble term", since it destroys an electron at the site $j$ and creates it at the site $j+1$ of the ring, but also destroys another electron at the same site $j+1$ displacing it to the site $j$. In other words, such term restricts the electronic motion between two neighboring sites of the ring. On the other hand, the second term, which is proportional to $c_{j\sigma}^{\dag}c_{j-1\sigma'}^{\dag}c_{j-2\sigma'}^{\null}c_{j-1\sigma}^{\null}$, involves two neighboring sites and \textit{favors the electron delocalization}.

Combining Eq.(\ref{eq:Weff2}) with Eq.(\ref{eq:H0}), we obtain the following extended Hubbard model for the $\pi$-electrons: 
\begin{widetext}

\begin{align}
\hat{H}=&-t\sum\limits_{j=1}^{N}\sum\limits_{\sigma}\left(c_{j\sigma}^{\dag}c_{j+1\sigma}^{\null}+\text{ h.c.}\right)+U\sum\limits_{j=1}^{N}\hat{n}_{j\uparrow}\hat{n}_{j\downarrow}\nonumber\\
-&\lambda_{N}\left(\frac{U}{t}\right)^2\sum\limits_{j=1}^{N}\sum\limits_{\sigma,\sigma'}\left[\left(c_{j\sigma}^{\dag}c_{j+1\sigma'}^{\dag}c_{j\sigma'}^{\null}c_{j+1\sigma}^{\null}+\text{h.c.}\right)+\left(c_{j\sigma}^{\dag}c_{j-1\sigma'}^{\dag}c_{j-2\sigma'}^{\null}c_{j-1\sigma}^{\null}+\text{h.c.}\right)\right] \text{,}
\label{eq:H}
\end{align}
\\
\end{widetext}
where we define the coupling constant 
\begin{equation}
\lambda_{N}\equiv 2\frac{t^4}{\Lambda^{3}}\mathcal{O}_0 \left(1-\cos^2(2\pi/N)\right) \text{ .}
\label{eq:lambda}
\end{equation}

\noindent Recall that $\Lambda>0$ is the energy scale of the separation between the ground and the first excited states of the $\sigma$-electrons, which we approximate by a constant, i.e., independent of the $\pi$-electrons configuration. Importantly, although the coupling constant carries information about the direction of the in-plane bonds though the term $\cos(2\pi/N)$, it is the relation between the parameters $t$ and $\Lambda$ that sets the energy scale of the coupling $\lambda_{N}$. Therefore, hereafter we set $2\mathcal{O}_0/(1-\cos^2(2\pi/N))\sim 1$, making the coupling constant site independet, $\lambda\equiv t^{4}/\Lambda^{3}<1$. At this point, it becomes clear that $\lambda$ is indeed the parameter that controls the validity of our generalized Born-Oppenheimer approach to this problem. If $t/\Lambda \ll 1$, the effective momentum-momentum coupling is negligible, and we end up with the standard Hubbard model where the bonding electrons are frozen at the chemical bonds and only dress the ionic potential experienced by the itinerant electrons. In other words,  Eq.(\ref{eq:H}) shows that, in systems where the bonding electrons behave as if they were in a covalent bond, the first correction to the motion of the itinerant electrons due to the remaining electrons of the system takes place by the generation of an effective inter-electronic momentum-momentum between the former. 

\begin{figure*}[t!]
\centering
\includegraphics[width=1\linewidth]{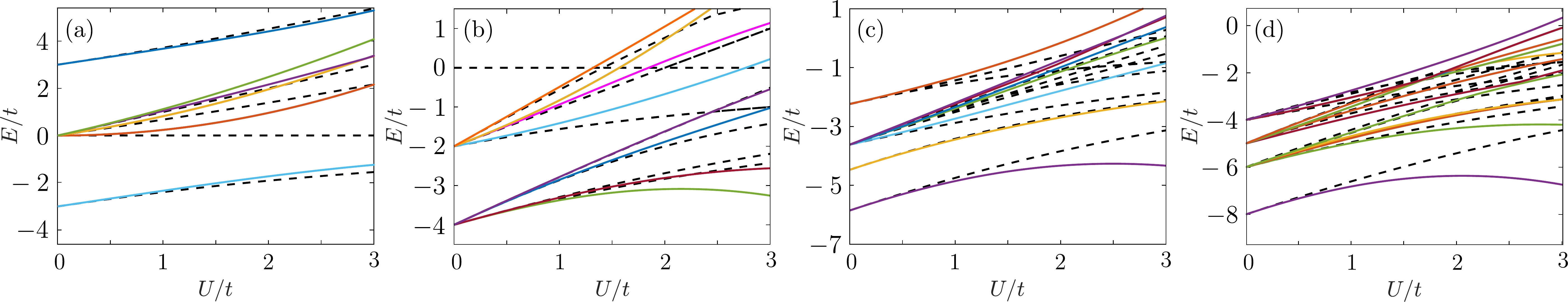}
\caption{\textbf{Energy spectrum of the extended Hubbard Hamiltonian.} The panels show the energy levels of Eq.(\ref{eq:H}), as a function of $U/t$ for rings with (a) $N=3$ sites, (b) $N=4$ sites, (c) $N=5$ sites, and (d) $N=6$ sites at the half-filling regime ($N=N_e$) and with $\lambda/t=0.1$. In the panels (b) to (d), only a few of the low-lying energy levels are plotted. The dashed lines correspond $\lambda=0$, i.e., the spectrum of the standard Hubbard model defined in Eq.(\ref{eq:H0}).}
\label{fig:newspectrum}
\end{figure*}

As Eq. (\ref{eq:H}) is the central result of this work, we now investigate some of its physical consequences. In particular, we focus on the energy spectrum of the rings. Fig.\ref{fig:newspectrum} highlights the impact of the effective interaction (\ref{eq:Weff2}) into the energy spectrum of the rings. The solid lines correspond to the lowest energy levels of a ring with (a) $N=3$, (b) $N=4$, (c) $N=5$ and $N=6$ sites at the half-filling regime ($N=N_e$) obtained through exact diagonalization of Eq. (\ref{eq:H}) with $\lambda/t=0.1$. The dashed lines in each panel (shown here for comparison purposes) refer to the energy level of the standard Hubbard model Eq.(\ref{eq:H0}). We verify that the effective momentum-momentum interaction between the $\pi$-electrons [Eq.(\ref{eq:Weff2})] always reduces the energy of the ground state of the rings, while there is no systematic behavior concerning their excited states.

The non-monotonic behavior of the ground state energy ($E_0$) as a function of $U/t$ results from the competition between the on-site Coulomb repulsion and the momentum-momentum effective coupling. Initially, $E_0$ tends to increase with $U$, since the on-site Coulomb repulsion favors localization. However, for large enough $U$, the effects of the attractive momentum-momentum interaction dominate and $E_0$ decreases. The larger $\lambda/t$ is, the sooner such a crossover takes place. We emphasize, however,  that for the perturbation scheme and approximations we used in the derivation of Eq.(\ref{eq:Weff2}) to be valid, we need to have  $\lambda (U/t)^2 < t$, or equivalently, $(t/\Lambda)^3  (U/t)^2 = (t/\Lambda) (U/\Lambda)^2 < 1$. This condition implies that, no matter what the ratio $U/t$ is, the spectra of Fig.\ref{fig:newspectrum}  make perfect sense for either $t<U<\Lambda$ or $U<t<\Lambda$, which fulfil the previously assumed conditions $t/\Lambda <1$ and $U/\Lambda<1$. Actually, for a fixed ratio $t/\Lambda$, the restriction on the maximum value for $U/t$ obeys the condition $U/t<(\Lambda/t)^{3/2}$.

In realistic situations we believe that what our approximation really does is provide an extension of the Hubbard model which is suitable for applications to broader band correlated systems, but more precise conclusions on this issue would require a more careful analysis of the Wigner-Brillouin perturbative method in Sec. \ref{BO_Approx}. From our present approach, and knowledge of correlated electronic systems, we expect this term to be innocuous in the non-interacting H\"uckel limit ($U/t << 1$) and completely overwhelmed by localization effects in very narrow band systems ($U/t >>1$). Once again, conclusions of this sort strongly depend, in particular, on the ratio $U/\Lambda$ which ultimately limits the extension of each many-body energy level as a function of $U/t$ in Fig.\ref{fig:newspectrum}.

 Although the spectra of Fig.\ref{fig:newspectrum} provide only a glimpse about the physics of each  particular example, there is a clear signature that even the ground state energies of those half filled systems  may be modified by the presence of the momentum-momentum coupling in the Hubbard Hamiltonian. Moreover, this lowering of the ground state energy does not happen only in those cases where they are known to be degenerate, such as in the $N=3,4,$ or $5$ cases, but also in our prototype of the benzene molecule, $N=6$, which has a non-degenerate ground state. Despite the physics of process (b) of Fig.\ref{fig:processes} suggests that a current-carrying state would have its energy lowered by this term, it is not plausible that such a state could be spontaneously generated in a finite dimensional system. Nevertheless, in order to test whether this reasoning makes any sense at all we can apply an external magnetic field perpendicular to the ring's plane which generates a persistent current of the itinerant electrons and analyze the outcome of our model in this case.

In a separate contribution \cite{paper2} we have shown that in the presence of the  external field, the effective momentum-momentum interaction Eq.(\ref{eq:Weff2}) amplifies the persistent electric current in the system's ground state leading to a strong enhancement of the component of its magnetic susceptibility tensor parallel to the field. Consequently, the effective momentum-momentum interaction could account for the well-known unusual magnetic anisotropy of aromatic molecules \cite{Krishhnan,Pauling}, whose microscopic origin remains a topic of debate in the literature \cite{Lazzeretti}.

Actually, the presence of the momentum-momentum interaction opens up a whole avenue for investigating how it affects different properties of the realistic systems our minimal models of Fig.\ref{fig:newspectrum} represent. For example, we could study how it influences the tendency of formation of the cyclopropenium cation - $\textrm{C}_3\textrm{H}_3^+$-, or the cyclopentadienide anion - $\textrm{C}_5\textrm{H}_5^-$-. These systems belong to a class of problems away from half-filling, where the effects of the momentum-momentum coupling are more dramatic, and do deserve to be studied separately. Our model can also play some role in the dimerization of the cyclobutadiene molecule - $\textrm{C}_4\textrm{H}_4$- where it could enhance the well-known contribution of the Jahn-Teller effect, if lattice distortions are allowed.  In these three cases, we should expect  process (a) of Fig.\ref{fig:processes} to be the most relevant one, since it favors the double electronic hopping along a given bond.



\section{Conclusions} \label{C}

Although the $\sigma$-electrons are more localized than the $\pi$-electrons, they can undergo local excitations in the $\sigma$-bonds. These excitations modify the electron charge density in the bonds, and, therefore renormalize the periodic potential felt by the $\pi$-electrons. We showed that such virtual excitations of the $\sigma$-electrons mediate an attractive momentum-momentum effective interaction between the $\pi$-electrons, which is described by $\hat{W}_{eff}$ defined in Eq.(\ref{eq:Weff_final}). This interaction bears some similarities with the  Biot-Savart kind of interaction derived by Pines and Bohm \cite{PinesI,PinesII,PinesIII}. 

Motivated by a natural energy scale separation between the $\sigma$ and $\pi$-electrons, we could decouple their degrees of freedom through the employment of a wave function ansatz similar to that used in the standard Born-Oppenheimer approximation. The resulting single-band problem was treated within first-order perturbation theory generating the above-mentioned momentum-momentum-effective interaction. 

There is a myriad of physical problems where the consequences of the existence of this kind of interaction could be tested. In this work, we address only its effect on the lowest-lying energy levels of some small rings, and compared it to the results obtained by the exact diagonalization of standard Hubbard Hamiltonian as applied to the same systems. We showed that the effective momentum-momentum interaction competes with the on-site Coulomb repulsion, dominating as $U/t$ grows. When this happens, the system's ground state energy is reduced significantly. The effective interaction (\ref{eq:Weff2}) also has important implications on the magnetic properties of the rings. In particular, it leads to a substantial enhancement of the magnetic response of the system and, therefore, might be linked to the microscopic origin of the magnetic anisotropy observed in aromatic molecules. This topic has recently been addressed elsewhere \cite{paper2}. 

We emphasize that the results we presented here relies on the exact diagonalization of small systems. In order to improve our physical understanding of the effects Eq.(\ref{eq:Weff2}), it would be interesting to implement other approximation methods such as, for example, mean-field theory. This is something we are aiming at, and hoping to present in future contributions.

Another important question one should raise at this point is why the effects of this term have never been observed so far. Well, in principle, we cannot say that they have never been observed. As we have already stressed, the importance of this term lies in a region of the parameter space at which more realistic systems should not be treated by simplified many-body models. In our particular case, we are dealing with the moderate $U$ regime, which is known to be the one to provide us with a poor physical picture when modeling realistic systems within the Hubbard scheme. The regimes $U\ll t$ or $U\gg t$ (narrow band systems) are much more appropriate to describe specific concrete situations. Therefore, we think that results arising from the extended Hubbard Hamiltonian should be compared, for example, with those obtained by the density functional theory, where one can simulate the dynamics of all kinds of electrons present in the system without drastic approximations. In this sort of comparison, the effects of the new term might be buried among others we could not reach in our approximation. Nevertheless, we hope this helps us select the conditions at which the momentum-momentum coupling generates the main contribution to the observed behavior. 

Finally, a few words about the generality of our result. Despite our conclusions has been obtained from a model where we reincorporate the dynamics of the bonding electrons into that of the itinerant electrons, the bottom line of our approach is the fact that we are dealing with two sets of interacting electrons whose typical energy scale is quite different from that of the former. Whereas in the present case, we have dealt with $\sigma$ and $\pi$-electrons, Bohm and Pines \cite{PinesI,PinesII,PinesIII} treated electron-hole pairs interacting with the long-wavelength longitudinal excitations (plasmons) of the same electron gas. Although the latter authors did not recognize the suitability of the adiabatic approximation, and perturbative corrections to it, to their system, we think we have successfully done it to ours. 

Furthermore, there is nothing in our method which prevents us from applying it to more general systems once we identify the existence of the basic ingredients necessary for the implementation of the generalized Born-Oppenheimer approximation, and consequently the generation of an effective momentum-momentum interaction between the components of the electronic system of interest. It should be, at least instructive, to study the possible effects of this term in $2D$ electronic systems such as graphene sheets or Cu-O planes in high-$T_c$ materials.  Not less important would be to study the influence of such a term in situations where the Hubbard model is known to be described as a Fermi liquid theory. In this case we could either find corrections to the Fermi liquid parameters or establish the conditions under which it would give rise to non-Fermi liquid effects.

\begin{acknowledgments}
We acknowledge S\~{a}o Paulo Research Foundation (FAPESP) and Conselho Nacional de Desenvolvimento Cient\'ifico e Tecnol\'ogico (CNPq) for the financial support. TVT and GMM were supported by FAPESP under the projects 2015/21349-7 and 2016/13517-0, respectively, whereas AOC was supported by CNPq under the project 302420/2015-0.
\end{acknowledgments}

\appendix

 \section{Effective interaction in first quantization}\label{first}

The first thing we need to do is return to Eqs.(\ref{eq:Anu})-(\ref{eq:gmunu}) and find an approximate expression for it. Let us start studying $\mathbf{g}_{\nu\mu}^{(j)}(\mathbf{R})$. Taking the gradient of Eq.(\ref{eq:sigma_Sch}) with respect to the position of the $j$-th $\pi$-electron, multiplying the result on the left by $\varphi_{\nu}^{*}(\mathbf{r},\mathbf{R})$, and integrating over the positions of the $\sigma$-electrons, we obtain
\begin{widetext}
\begin{equation}
\int d\mathbf{r}\,\varphi_{\nu}^{*}(\mathbf{r},\mathbf{R})\boldsymbol{\nabla}_{j}\varphi_{\mu}(\mathbf{r},\mathbf{R})=\frac{1}{\lambda_{\mu}(\mathbf{R})-\lambda_{\nu}(\mathbf{R})}\int d\mathbf{r}\varphi_{\nu}^{*}(\mathbf{r},\mathbf{R})\left(\boldsymbol{\nabla}_{j}\mathcal{H}_{b}\right)\varphi_{\mu}(\mathbf{r},\mathbf{R}) \text{ .}
\label{eq:int_step}
\end{equation}
\end{widetext}

\noindent Recall that $\mathbf{R}$ is merely an external parameter for $\mathcal{H}_{b}$, and it appears only in the Coulomb repulsion term (see Eq.(\ref{eq:Hb})). Therefore, it follows that 
\begin{equation}
\boldsymbol{\nabla}_{j}\mathcal{H}_b(\mathbf{R})=e^{2}\sum\limits_{\alpha=1}^{N_{e}^{(\sigma)}}\frac{\mathbf{r}_{\alpha}-\mathbf{R}_{j}}{\left|\mathbf{r}_{\alpha}-\mathbf{R}_{j}\right|^{3}} \text{ .}
\label{eq:grad}
\end{equation}

Our task now is to evaluate the integral over the positions of the $\sigma$-electrons. Since we have a term $|\mathbf{r}_{\alpha}-\mathbf{R}_{j}|^3$ in the denominator of the integrand, the $\sigma$-electrons which are closer to the $j$-th $\pi$-electron are those which give the largest contribution to the right-hand side of Eq.(\ref{eq:int_step}). Furthermore, we have two $\sigma$-electrons per bond. Consequently, for each $\pi$-electron localized at a given ring site, there are four nearest neighbors $\sigma$-electrons, here labeled by $1$ to $4$ for simplicity, that dominates the sum in Eq.(\ref{eq:grad}), which we can approximate as
\begin{widetext}
\begin{align}
&\int d\mathbf{r}\,\varphi_{\nu}^{*}(\mathbf{r},\mathbf{R})\left(\boldsymbol{\nabla}_{j}\mathcal{H}_{b}\right)\varphi_{\mu}(\mathbf{r},\mathbf{R})\approx
e^{2}\hat{\mathbf{d}}_{j}^{(L)}\int d\mathbf{r}\,\varphi_{\nu}^{*}(\mathbf{r},\mathbf{R})\left(\frac{1}{\left|\mathbf{r}_{1}-\mathbf{R}_{j}\right|^2}+\frac{1}{\left|\mathbf{r}_{2}-\mathbf{R}_{j}\right|^2}\right)\varphi_{\mu}(\mathbf{r},\mathbf{R})+ \nonumber \\[0.2cm]
&+ e^{2}\hat{\mathbf{d}}_{j}^{(R)}\int d\mathbf{r}\,\varphi_{\nu}^{*}(\mathbf{r},\mathbf{R})\left(\frac{1}{\left|\mathbf{r}_{3}-\mathbf{R}_{j}\right|^2}+\frac{1}{\left|\mathbf{r}_{4}-\mathbf{R}_{j}\right|^2}\right)\varphi_{\mu}(\mathbf{r},\mathbf{R}) \text{ .}
\label{eq:ap}
\end{align}
\end{widetext} 

\noindent We define
\begin{equation}
\hat{\mathbf{d}}_{j}^{(R)}\equiv \frac{1}{a}\left(\mbox{\boldmath$\mathcal{R}$} _{j+1}-\mbox{\boldmath$\mathcal{R}$} _{j}\right)
\label{eq:d}
\end{equation}

\noindent as the versor in the direction of the \textit{right} $\sigma$-bond, between the sites $j$ and $j+1$. Recall that $\mbox{\boldmath$\mathcal{R}$} _{j}$ is the position of site $j$ defined in Eq.(\ref{eq:R1}). 

Concerning the remaining integrals on the right-hand side of Eq.(\ref{eq:ap}), if we had $\left|\mathbf{r}_{\alpha}-\mathbf{R}_{j}\right|$ in the denominator, they would be of the order of the on-site Coulomb repulsion between $\pi$-electrons and $\sigma$-electrons ($\tilde{U}$) which we assume to be of the same order as the on-site repulsion ($U$) between the $\pi$-electrons. Moreover, since  $\left|\mathbf{r}_{\alpha}-\mathbf{R}_{j}\right|$ is of the order of the lattice spacing $a$ we can roughly estimate for $\mu\neq\nu=0,1$, 
\begin{equation}
e^{2}\int d\mathbf{r}\,\varphi_{\nu}^*(\mathbf{r},\mathbf{R})\frac{1}{\left|\mathbf{r}_{\alpha}-\mathbf{R}_{j}\right|^2}\varphi_{\mu}(\mathbf{r},\mathbf{R})\sim \frac{U}{a} \text{ ,}
\end{equation}  

\noindent with $\alpha=1,2,3,4$. Therefore, Eq.(\ref{eq:ap}) results in
\begin{align}
\int &d\mathbf{r}\,\varphi_{\nu}^*(\mathbf{r},\mathbf{R})\left(\boldsymbol{\nabla}_{j}\mathcal{H}_{b}\right)\varphi_{\mu}(\mathbf{r},\mathbf{R})\approx\nonumber\\
&\approx 2\frac{U}{a}\left(\hat{\mathbf{d}}_{j}^{(R)}-\hat{\mathbf{d}}_{j-1}^{(R)}\right)\text{.}
\label{eq:result}
\end{align}

Substituting Eq.(\ref{eq:result}) into Eq.(\ref{eq:int_step}) and comparing it with (\ref{eq:gmunu}) we readily identify
\begin{equation}
\mathbf{g}_{\nu\mu}^{(j)}(\mathbf{R})\approx-\frac{i\hbar}{am}\frac{2U}{\lambda_{\mu}(\mathbf{R})-\lambda_{\nu}(\mathbf{R})}\left(\hat{\mathbf{d}}_{j}^{(R)}-\hat{\mathbf{d}}_{j-1}^{(R)}\right) \text{.}
\label{eq:approx_g}
\end{equation} 

Following similar steps for $f_{\nu\mu}(\mathbf{R})$, we find that it is of the order of $(U/\Lambda)^2$, and therefore can be neglected as long as $(U/\Lambda)^2\ll 1$, which we assume to be the case here. Moreover, in contrast to $\mathbf{g}_{\nu\mu}^{(j)}(\mathbf{R})$, $f_{\nu\mu}(\mathbf{R})$ does not involve the $\pi$-electrons momenta and, therefore, when included in Eq.(\ref{eq:potential_eff}) gives rise, in first order perturbation theory, to a one-body term that can be incorporated in the hopping parameter. Therefore, from Eq.(\ref{eq:approx_g}), and as in the previous section, assuming $\lambda_{1}(\mathbf{R})-\lambda_{0}(\mathbf{R})\approx \Lambda>0$ (constant), we can approximate $\mathcal{A}_{01}$ and $\mathcal{A}_{10}$ by Eqs.
~(\ref{eq:a01}) and (\ref{eq:a10}).

\section{Matrix element of $\mathbf{P}$ in the site basis}\label{P}

Here, we derive Eq.(\ref{eq:Pij}). To calculate the matrix element of the $\pi$-electrons momentum, we use the electron \textit{velocity operator} $\mathbf{V}\equiv \mathbf{P}/m$, which is related with the system's single-particle Hamiltonian ($h$) though the commutator
\begin{equation}
	\mathbf{V}=\frac{1}{i\hbar}\left [ \mathbf{R},h \right ] \text{ .}
	\label{eqC:commutator}
\end{equation}

\noindent Here,
\begin{equation}
h(\mathbf{r})\equiv\frac{\mathbf{P}^2}{2m}+V_{c}(\mathbf{r})=-\frac{\hbar^2}{2m}\boldsymbol{\nabla}^{2}+V_{c}(\mathbf{r}) \text{ ,}\label{eqA:h}
\end{equation}

\noindent and we denote by $\mathbf{R}$ the electron position operator. 

Evaluating the matrix element of Eq.(\ref{eqC:commutator}) in the single-particle Wannier wave functions $\varphi_{j}(\mathbf{r})=\left\langle \left.\mathbf{r}\right|j\right\rangle$, we obtain
\begin{eqnarray}
\frac{1}{m}\left \langle j_1 \left |\mathbf{P}  \right |j_2 \right \rangle&=&\left \langle j_1\left | \frac{1}{i\hbar}\left [ \mathbf{\mathbf{R}},h \right ] \right |j_2\right \rangle\nonumber\\
&=&\frac{1}{i\hbar}\left ( \left \langle j_1\left | \mathbf{R}h \right | j_2\right \rangle - \left \langle j_1\left | h\mathbf{R} \right | j_2\right \rangle \right ).
\label{eqC:Pv1}
\end{eqnarray}

\noindent Now, inserting the closure relation
\begin{equation}
\mathbb{1}=\sum\limits_{j=1}^{N}\left|j\right\rangle\left\langle j\right|
\end{equation}

\noindent between the $\mathbf{R}$ and $h$ operators on the right-hand side of Eq.(\ref{eqC:Pv1}) and approximating the position matrix elements as 
\begin{equation}
\left\langle j_1\left|\mathbf{R}\right|j_2\right\rangle\approx \mbox{\boldmath$\mathcal{R}$}_{j_2}\left\langle j_1\left|\right. j_2\right\rangle=\mbox{\boldmath$\mathcal{R}$}_{j_2}\delta_{j_1,j_2} \text{ ,}
\end{equation}

\noindent which is justified by the fact that the Wannier function $\varphi_{j}(\mathbf{r})$ is localized about the $j$-th site of the ring, whose position we denote by $\mbox{\boldmath$\mathcal{R}$}_{j}$, we readily find 
\begin{equation}
\frac{1}{m}\left \langle j_1 \left |\mathbf{P}  \right |j_2 \right \rangle=\frac{1}{i\hbar}\left(\mbox{\boldmath$\mathcal{R}$}_{j_1}-\mbox{\boldmath$\mathcal{R}$}_{j_2}\right)\left\langle j_{1}\left|h\right| j_2 \right\rangle \text{ .}
\end{equation}

Finally, recalling that $\left\langle j_{1}\left|h\right| j_2 \right\rangle=t_{j_1,j_2}$ gives the hopping between the sites $j_{1}$ and $j_{2}$, which, in the nearest-neighbor approximation, simplifies to 
\begin{equation}
\left\langle j_{1}\left|h\right| j_2 \right\rangle\approx -t\delta_{j_{2},j_{1}\pm 1} \text{ ,}
\end{equation}

\noindent we obtain Eq.(\ref{eq:Pij}):
\begin{equation}
\left \langle j_1 \left |\mathbf{P}  \right |j_2 \right\rangle =-\frac{mt}{i\hbar}\left(\mbox{\boldmath$\mathcal{R}$}_{j_1}-\mbox{\boldmath$\mathcal{R}$}_{j_2}\right)\delta_{j_{2},j_{1}\pm 1}.
\end{equation}

\section{Evaluation of $w_{ijkl}$ }\label{w}
In this Appendix we show the intermediate steps between Eqs.(\ref{eq:w3}) and (\ref{eq:w5}). Our main assumption here involves the matrix elements of the tensor $\overleftrightarrow{T}$. Similarly to what is done with the Coulomb repulsion matrix elements in the standard Hubbard model \cite{Hubbard1}, we consider that its leading contributions come from the on-site terms:
\begin{equation}
 \left\langle ij\left|\overleftrightarrow{T}\right|lk\right\rangle\approx\overleftrightarrow{T_{i}}\, \delta_{j,i}\delta_{k,i}\delta_{l,i} \text{ ,}
\end{equation} 
\noindent where we define $\overleftrightarrow{T_{i}}\equiv \left\langle ii\left|\overleftrightarrow{T}\right|ii\right\rangle$.  As it will become clearer soon, $\overleftrightarrow{T_{i}}$ is site-dependent.

Returning to the definition of $\overleftrightarrow{T}$ in Eq.(\ref{eq:T}), we can write
\begin{equation}
\overleftrightarrow{T_{i}}=\int\!\!\int d\mathbf{r}\,d\mathbf{r'}\,\varphi_{i}^{*}(\mathbf{r})\varphi_{i}^{*}(\mathbf{r'})\hat{\mathbf{n}}\mathcal{O}(\mathbf{r},\mathbf{r'})\hat{\mathbf{n}}'\varphi_{i}(\mathbf{r'})\varphi_{i}(\mathbf{r}) \textit{ ,}
\end{equation}
with $\hat{\bf{n}}\equiv\mathbf{r}/r$ and $\hat{\bf{n}}'\equiv\mathbf{r'}/r'$. Moreover, since the Wannier wave functions are localized at the ring's sites, 
\begin{align}
\overleftrightarrow{T_{i}}&=\hat{\mbox{\boldmath$\mathcal{R}$} _{i}}\left[\int\!\!\int d\mathbf{r}\mathbf{r'}\varphi_{i}^{*}(\mathbf{r})\varphi_{i}^{*}(\mathbf{r'})\mathcal{O}(\mathbf{r},\mathbf{r'})\varphi_{i}^{\null}(\mathbf{r'})\varphi_{i}^{\null}(\mathbf{r})\right]\hat{\mbox{\boldmath$\mathcal{R}$} _{i}}\nonumber\\[0.2cm]
&=\hat{\mbox{\boldmath$\mathcal{R}$} _{i}}\left\langle i i\left|\mathcal{O}\right| i i \right\rangle \hat{\mbox{\boldmath$\mathcal{R}$} _{i}}
\label{eq:Ti}
\end{align}  

\noindent with $\hat{\mbox{\boldmath$\mathcal{R}$} _{i}}=\mbox{\boldmath$\mathcal{R}$} _{i}/\left|\mbox{\boldmath$\mathcal{R}$} _{i}\right|$ denoting the unitary vector in the direction of the position of site $j$. Assuming, for simplicity, that the matrix element of $\mathcal{O}(\mathbf{r},\mathbf{r})$ is homogeneous, i.e. $\left\langle i i \left|\mathcal{O}\right| i i \right\rangle=\mathcal{O}_0$ is \textit{site-independent}, and that $\mathcal{O}_0$ is a scalar of order one, we rewrite Eq.(\ref{eq:w3}) as
\begin{widetext}
\begin{align}
    w_{ijkl}\approx-&\left(\frac{mt}{\hbar}\right)^2\mathcal{O}_0\times \left[\left(\mbox{\boldmath$\mathcal{R}$} _{i}-\mbox{\boldmath$\mathcal{R}$} _{i+1}\right)\cdot\hat{\mbox{\boldmath$\mathcal{R}$} }_{i+1}\,\left(\mbox{\boldmath$\mathcal{R}$} _{i+1}-\mbox{\boldmath$\mathcal{R}$} _{i}\right)\cdot \hat{\mbox{\boldmath$\mathcal{R}$} }_{i+1}\,\delta_{j,i+1}\delta_{l,i+1}\delta_{k,i}\nonumber\right.\\
    +&\left(\mbox{\boldmath$\mathcal{R}$} _{i}-\mbox{\boldmath$\mathcal{R}$} _{i+1}\right)\cdot \hat{\mbox{\boldmath$\mathcal{R}$} }_{i+1}\,\left(\mbox{\boldmath$\mathcal{R}$} _{i+1}-\mbox{\boldmath$\mathcal{R}$} _{i+2}\right)\cdot \hat{\mbox{\boldmath$\mathcal{R}$} }_{i+1}\,\delta_{j,i+1}\delta_{l,i+1}\delta_{k,i+2}\nonumber\\
    +&\left(\mbox{\boldmath$\mathcal{R}$} _{i}-\mbox{\boldmath$\mathcal{R}$} _{i-1}\right)\cdot \hat{\mbox{\boldmath$\mathcal{R}$} }_{i-1}\,\left(\mbox{\boldmath$\mathcal{R}$} _{i-1}-\mbox{\boldmath$\mathcal{R}$} _{i-2}\right)\cdot \hat{\mbox{\boldmath$\mathcal{R}$} }_{i-1}\,\delta_{j,i-1}\delta_{l,i-1}\delta_{k,i-2}\nonumber\\
    +&\left.\left(\mbox{\boldmath$\mathcal{R}$} _{i}-\mbox{\boldmath$\mathcal{R}$} _{i-1}\right)\cdot\hat{\mbox{\boldmath$\mathcal{R}$} }_{i-1}\,\left(\mbox{\boldmath$\mathcal{R}$} _{i-1}-\mbox{\boldmath$\mathcal{R}$} _{i}\right)\cdot \hat{\mbox{\boldmath$\mathcal{R}$} }_{i-1}\delta_{j,i-1}\delta_{l,i-1}\delta_{k,i}\right]\!\!\text{.}
    \label{eq:w4}
\end{align}
\end{widetext}

Moreover, using that, for a $N$-site ring
\begin{align}
&\mbox{\boldmath$\mathcal{R}$} _{i}\cdot\hat{\mbox{\boldmath$\mathcal{R}$} }_{i}=\frac{a}{\sqrt{2\left(1-\cos(2\pi/N)\right)}} \text{ ,}\\
&\mbox{\boldmath$\mathcal{R}$} _{i\pm 1}\cdot\hat{\mbox{\boldmath$\mathcal{R}$} }_{i}=\frac{a}{\sqrt{2\left(1-\cos(2\pi/N)\right)}}\cos\left(\frac{2\pi}{N}\right) \text{ ,} 
\end{align}
\noindent where $a$ is the system's lattice spacing, Eq.(\ref{eq:w4}) readily simplifies to (\ref{eq:w5}).


\end{document}